\documentclass[aps,pra,twocolumn,showpacs,floatfix]{revtex4}
\usepackage{epsfig}
\usepackage{graphicx}
\usepackage{dcolumn}
\usepackage{amsthm,amsmath}

\begin{document}
\title{Dispersion $C_3$ coefficients for the alkali-metal atoms interacting with a graphene layer and with a carbon nanotube}
\author{$^a$Bindiya Arora \footnote{Email: arorabindiya@gmail.com}, $^a$Harjeet Kaur and $^b$B. K. Sahoo  \footnote{Email: bijaya@prl.res.in}}
\affiliation{$^a$Department of Physics, Guru Nanak Dev University, Amritsar, Punjab-143005, India,\\
$^b$Theoretical Physics Division, Physical Research Laboratory, Navrangpura, Ahmedabad-380009, India}
\date{Received date; Accepted date}
 
\begin{abstract}
We evaluate separation dependent van der Waal dispersion ($C_3$) coefficients for the interactions of the Li, Na, K and Rb alkali atoms 
with a graphene layer and with a single walled carbon nanotube (CNT) using the hydrodynamic and Dirac models. The results from both the models 
are evaluated using accurate values of the dynamic polarizabilities of the above atoms. Accountability of these accurate values of dynamical 
polarizabilities of the alkali atoms in determination of the above $C_3$ coefficients are accentuated by comparing them with the coefficients 
evaluated using the dynamic dipole polarizabilities estimated from the single oscillator approximation which are typically employed in the 
earlier calculations. For practical description of the atom-surface interaction potentials the radial dependent $C_3$ coefficients are 
given for a wide range of separation distances between the ground states of the considered atoms and the wall surfaces and also for
different values of nanotube radii. The coefficients for the graphene layer are fit to a logistic function dependent on the separation distance.
For CNT, we have carried out a paraboloid kind of fit dependent on both the separation distances and radii of the CNT. These fitted functions,
with the list of fitting parameters, can be used to extrapolate the interaction potentials between the considered alkali atoms and the graphene 
layer or CNT surface conveniently at the given level of accuracy.
\end{abstract}

\maketitle
 
\section{Introduction}\label{sec1}
In the past decade considerable amount of attention have been drawn both towards the experimental and theoretical studies of the internal scattering
of atoms with a graphene layer and with various carbon nanostructures \cite{agra, fermani,friedrich}. Owing to the fact that these nanostructures are endowed with
exceptional electronic, optical, mechanical, thermal, and magnetic properties that are of vested interest to the modern communication engineering 
technologies~\cite{intro1,intro2}, their applications are in huge demand both in the scientific and industrial laboratories. Graphene, in particular,
manifest unique properties incurring its honey comb-lattice structure that could maximize the interaction of atom on the layer. In fact, the 
knowledge of the atom-graphene interactions has been very useful in the construction of the hydrogen storage devices \cite{intro3,intro4,intro5} and
also plays an important role in understanding different physical, chemical and biological processes~\cite{intro6,intro7,intro8,intro9}. Moreover, these interactions are 
connected to the phenomenon of quantum reflections whose studies are of special interest today to many experimentalists and theoreticians for 
explaining their exact behavior ~\cite{dirac-hydro,qr1,qr2,qr3}. In addition to this, gaining insights of atom-graphene contacts are crucial in the 
development of graphene based electronics. In particular, metals adsorbed on graphene can form different types of structures and can change 
graphene's electronic behavior instigating towards observation of interesting physical phenomenon~\cite{intro10,intro12,intro13,intro14}. Among the 
metals that can be adsorbed on graphene, study for the Li atom is particularly very interesting for the applications in the storage of hydrogen gas \cite{ataca,du}, 
improving efficiencies of Li-ion batteries \cite{konga,jian}, and making superconductors~\cite{castro,intro16}. K atoms have also been used to tune the 
electronic structures of the graphene bilayers~\cite{intro17,intro18}. A graphene layer rolled to make a carbon nanotube (CNT) has some very 
peculiar properties and gained special attention by the researchers world-wide \cite{cnt-1,cnt-2,cnt-3}. Interaction of the alkali atoms with the single walled 
CNTs have profound applications in the purification of the CNTs~\cite{cnt-pur}. Adsorbed alkali atoms have been demonstrated to act as chemical 
dopants on the CNTs and have been used to fabricate field effective transistors~\cite{cnt-fet}. 

Accurate experimental measurements of the $C_3$ coefficients of any atomic systems with a graphene layer and with a CNT is extremely difficult. 
A number of theoretical methods have been performed, particularly using the density functional theories \cite{dft1,dft2,dft3,dft4}, lower order 
many-body methods \cite{mbpt1}, Lifshitz theory ~\cite{babb,caride,lf1,lf2,nano1,dirac-hydro, dirac-hydro2} etc., to uncover the nature of 
interactions of carbon nanostructures with various materials. Klimchitskaya and co-worker have used Lifshitz theory to explain the interaction 
between graphene layer with different materials including metal plate~\cite{bordag,metal1,metal2}, conduction cylinder~\cite{cylinder}, atoms such 
as H~\cite{dirac-hydro}, Na, Rb, Cs~\cite{dirac-hydro,dirac-hydro2}, H$_2$ molecule~\cite{nano1}, He* ion~\cite{dirac-hydro,dirac-hydro2}, etc. 
Interaction of CNT with the H atom and H$_2$ molecule have also been explained by them in great detail~\cite{nano1}. However, these calculations
for graphene layer have been basically carried out by employing single oscillator model (SOM) for the estimation of the dynamic polarizabilities. In this 
paper, we verify the results by evaluating them for the alkali atoms with their accurate values of the dynamic polarizabilities and emphasis on 
the need to use such accurate values by comparing our results with the SOM results. Moreover, the interactions between CNT and alkali metal-atoms remain to 
be investigated thoroughly. In view of the fact that these interactions play crucial roles in a number of applications and keeping in mind
their vast experimental use, it would be expedient to carry out more accurate theoretical analysis of the interactions of the graphene layer and
CNT with the alkali atoms.

The interaction between an atom and a wall is usually modeled by calculating the interaction between the atom and its image charge (reflection) in 
the wall. The reflection coefficients required for such calculations are well described by the widely celebrated Lifshitz theory which expresses 
these quantities as the functions of the dynamic dielectric permittivity of the wall and of the dynamic dipole polarizabilities of the atoms 
\cite{harjeet1,lifshitzbook,klim,mahanty,pars}. Although accurate evaluation of the dynamic dipole polarizabilities in the atomic systems require 
sophisticated many-body methods, but their values for the alkali atoms, which are the utmost used atoms in the ultra-cold atomic experiments, are 
now known reliably at least with the sufficient precision at this stage of interest \cite{arora-sahoo1,arora-sahoo2}. In contrast, the dynamic 
dielectric permittivity values are generally known to insufficient accuracy in the material mediums due to their strenuous procedure of evaluation 
and could remain to be a tedious task for a long to determine them precisely. In particular, the nanostructures with thickness of the size of an atom such 
as the considered graphene layer and single walled CNT do not have well defined dielectric permittivity. This entails the need for adopting suitable
models to estimate the $C_3$ coefficients by introducing some effective parameters that can substitute the role of the dielectric permittivity of 
the wall in the Lifshitz theory. In this context, the two most popular models that are often employed in the theoretical determination of the 
dispersion coefficients are the hydrodynamic model \cite{barton,barton1,harjeet2,bordag1,blagov} and the Dirac model \cite{bordag2}. In this work, 
we intend to apply both the models and would like to compare the obtained results using accurate values of the dynamic dipole polarizabilities. 
In addition, we plan to present a very handy functional form of the radial dependent dispersion coefficients so that they can be easily 
derived for any arbitrary values of the atom-wall distance and the CNT radius for their convenient use in the practical applications.

This paper is organized as follows: In Sec. \ref{sec2}, we present the modified Lifshitz theory for the reflection coefficient on the graphene
layer and CNT in the hydrodynamic and Dirac model framework. This follows with a brief description of the method of calculations of the dynamic 
dipole polarizabilities in Sec. \ref{sec3} which are later used in the evaluation of the dispersion coefficients. Calculated results for the $C_3$ 
coefficients using accurate values of the dynamic polarizabilities and using the SOM model are given in Sec. \ref{sec4}. In the same section we present 
the dispersion coefficients for the graphene layer and CNT determined by employing both the 
hydrodynamic and Dirac models and compare them with the results obtained for an ideal conducting medium, Au, and SiO$_2$ wall which were reported
earlier. Unless stated explicitly, the results are given in atomic unit (a.u.) throughout the paper.

\section{Theory of the Dispersion Coefficient}\label{sec2}
The general form of the interaction potential energy in the configuration of a micro-particle and a material planar structure interacting at a distance $a$ is described by the Lifshitz theory which is expressed by \cite{lifshitz1,lifshitzbook}
{\small \begin{equation}
U(a)= -\frac{\alpha_{fs}^3}{2\pi}\int_0^{\infty}d\omega \omega^3\alpha(\iota\omega)\int_1^{\infty}d\xi e^{-2\alpha_{fs}\xi\omega a} H(\xi,\epsilon(\iota\omega)), \label{atwp}
\end{equation}}
where $\alpha_{fs}$ is the fine structure constant, $\epsilon(\omega)$ is the 
frequency dependent dielectric constant of wall material, $a$ is the separation 
distance between the atom and the surface and $\alpha(\iota\omega)$ is the dynamic
polarizability of the atom with imaginary argument. The function $H(\xi,\epsilon(\iota\omega))$
is given by
\begin{equation}
H(\xi,\epsilon)=(1-2\xi^2)\frac{\sqrt{\xi^2+\epsilon-1}-\epsilon\xi}{\sqrt{\xi^2+\epsilon-1}+\epsilon\xi} + \frac{\sqrt{\xi^2+\epsilon-1}-\xi}{\sqrt{\xi^2+\epsilon-1}+\xi}\nonumber
\end{equation}
with the Matsubara frequencies denoted by $\xi$.
 
For small separation distances, the above potential can be approximated to
\begin{equation}
U(a) \approx -\frac{C_3(a)}{a^3}, 
\label{u3}
\end{equation}
where $C_3$ is known as the dispersion coefficient for the corresponding atom-wall interaction. For a perfect conductor with $\epsilon(\omega) \rightarrow \infty$, 
we have
\begin{eqnarray}
C_3 &=& \frac{1}{4 \pi} \int_0^{\infty} d \omega \alpha(\iota \omega) \frac{\epsilon(\iota \omega)-1}{\epsilon(\iota \omega)+1} \nonumber \\
    &=& \frac{1}{4 \pi} \int_0^{\infty} d \omega \alpha(\iota \omega). 
\end{eqnarray}
The dispersion coefficient for a CNT with radius $R$ is expressed using the proximity force approximation (PFA) by \cite{som1,harjeet1,blagov} 
{\small
\begin{eqnarray}
C_3(a,R) &=& \frac{1}{16\pi}\sqrt{\frac{R}{R+a}}\int_0^{\infty} d\xi\alpha(\iota\xi)\int_{2a \alpha_{fs} \xi} dy y e^{-y} \nonumber \\
&&\left( y-\frac{a}{2(R+a)}\right) \left(2r_{\rm {TM}}-\frac{4a^2 \alpha_{fs}^2 \xi^2}{y^2}(r_{\rm{TM}}+r_{\rm{TE}})\right),\nonumber\\
\label{eq-c3cnt}
\end{eqnarray}}
where $r_{\rm{TM}}$ and $r_{\rm{TE}}$ are the reflection coefficients of the electromagnetic oscillations on CNT for the transverse magnetic 
and transverse electric polarizations of the electromagnetic field. 

It has been shown in Ref.~\cite{cylinder} that relative differences between the exact and PFA results
could be within 4\% for the condition $\frac{a}{R} < \frac{3}{5}$. For a thin single layer graphene with limit 
$R \rightarrow 0$, we can simplify the above expression to \cite{dirac-hydro}
\begin{eqnarray}
C_3(a)&=&\frac{1}{16\pi}\int_0^{\infty}d\xi\alpha(\iota\xi)\int_{2a\xi \alpha_{fs} }^{\infty}dye^{-y}y^2 \nonumber\\
&&\left(2r_{{TM}}-\frac{4a^2 \alpha_{fs}^2\xi^2}{y^2}(r_{{TM}}+r_{{TE}})\right).\label{eq-c3g}
\end{eqnarray} 
The separation distance dependent $C_3$ coefficients given above include both the retarded
and nonretarded interaction energies which are applicable up to the separation 
distances where the thermal effects are not significant (typically $\sim 1 \mu  m$)~\cite{casimir,babb}.

The most difficult part in the evaluation of the above expressions is to get the 
$r_{\rm{TM}}$ and $r_{\rm{TE}}$ reflection coefficients correctly.  
Two different models widely used to describe the electronic structure of graphene are the hydrodynamic model and the Dirac model.
Within the framework of hydrodynamic model, the reflection coefficients for a 
graphene layer or CNT are given by \cite{barton,barton1,harjeet2,bordag1,blagov}
\begin{eqnarray}
r_{\rm{TM}} &=& \frac{q \kappa }{q \kappa + \alpha_{fs}^2\xi^2} \nonumber\\
\text{and} \ \ \ r_{\rm{TE}}&=& -\frac{\kappa}{\kappa+q},\label{eq-h}
\end{eqnarray}
with the wave number of graphene sheet $\kappa=6.75 \times 10^{5} \ m^{-1}$ and $q=\frac{y}{2a}$.
In this model, graphene is considered as an infinitesimally thin positively 
charged sheet carrying a homogeneous fluid with some mass and negative charge 
densities.  The energy of the quasi-particles in graphene is quadratic with 
respect to their momenta. Therefore, this model works well at large energies and
fails at the low energies (where actual energy of the quasi-particles is linear function of 
momentum). This model is an approximate one and does not take into account the 
Dirac character of the charge carriers in graphene.

Within the framework of the Dirac model of the electronic structure of graphene, 
the quasi-particle fermion excitations in graphene are treated as massless Dirac
fermions moving with a Fermi velocity. It takes into account the 
properties of graphene which are valid at the low energies of the quasi-particles in graphene, specifically energies 
which are linear function of momentum. The explicit relations 
for the reflection coefficients considering the electronic structure of the graphene or CNT according to the Dirac model are given by \cite{bordag2}
\begin{eqnarray}
r_{\rm{TM}} &=& \frac{\alpha q \phi(\tilde{q})}{2{\tilde{q}}^2+\alpha q\phi(\tilde{q})} \nonumber\\
\text{and} \ \ \ r_{\rm{TE}}&=& -\frac{\alpha  \phi(\tilde{q})}{2 q+\alpha \phi(\tilde{q})},\label{eq-d}
\end{eqnarray}
where the function $\phi(\tilde{q})$ determines the polarization tensor in an external electromagnetic field in three dimension space-time coordinate and is give by \cite{bordag2} 
\begin{equation}
\phi(\tilde{q}) = 4 \left( \alpha_{fs} \Delta+\frac{\tilde{q}^2-4 \alpha_{fs}^2 \Delta^2}{2\tilde{q}}\rm{arctan} \left ( \frac{\tilde{q}}{2 \alpha_{fs} \Delta} \right ) \right),
\end{equation}
where $\Delta$ is known as the mass gap parameter. The exact value of 
$\Delta$ remains to be unknown however, its commonly accepted upper bound value quoted in the literature is 0.1 
eV ~\cite{castro,bordag2}. The parameter $\tilde{q}$ in the above equation is defined in terms of 
the Fermi velocity $v_f \sim 10^{6} \  m/s$ as
\begin{equation}
\tilde{q} = \left[ \frac{\alpha_{fs}^2 v_f^2 y^2}{4a^2}+\left(1-\alpha_{fs}^2 v_f^2\right) \alpha_{fs}^2 \xi^2 \right]^{1/2} .
\end{equation} 

In the next section, we shall briefly discuss the method of calculations for the 
dynamic polarizabilities which are required for evaluating $C_3$ coefficients 
as discussed above and would like to compare them with the results obtained 
considering the SOM results.

\section{Evaluation of the Dynamic Polarizabilities}\label{sec3}
The dynamic dipole polarizabilities of the ground state  $|\Psi_n\rangle$  of the alkali atoms
for the corresponding principal quantum number $n$ due to the direct current electric field
with the frequency $\omega$ are given by
\begin{eqnarray}
\alpha(\omega)&=& \sum_I \left [ \frac{ |\langle \Psi_n | D | \Psi_I\rangle |^2}{E_I - E_n + \omega}
               + \frac{ |\langle \Psi_n | D | \Psi_I\rangle |^2}{E_I - E_n - \omega} \right ] \nonumber \\
              &=& \frac{2}{3(2J_n+1)} \sum_I \frac{ (E_I-E_n) |\langle \Psi_n || D || \Psi_I\rangle |^2}{(E_I - E_n)^2 - \omega^2},
\label{pol1}
\end{eqnarray}
where $J_n=1/2$ is the total angular momentum of the corresponding ground state, sum over
$I$ represents all possible allowed intermediate states for the dipole transition, $E_S$ 
are the energies of the corresponding states and $\langle \Psi_n || D || \Psi_I\rangle$
is the E1 reduced matrix element of the dipole operator $D$ between the states $|\Psi_n\rangle$ and $|\Psi_I\rangle$.

Alternatively, the above polarizability expression can be expressed as
\begin{eqnarray}
\alpha(\omega)&=& \langle \Psi_n | D | \Psi_n^{(+)} \rangle 
               + \langle \Psi_n | D | \Psi_n^{(-)}\rangle ,
\label{pol2}
\end{eqnarray}
with 
\begin{eqnarray}
| \Psi_n^{(\pm)} \rangle &=& \sum_I  | \Psi_I\rangle \frac{\langle \Psi_I | D | \Psi_n\rangle}{E_I - E_n \pm \omega} 
\end{eqnarray}
which can be obtained for the Dirac-Coulomb (DC) Hamiltonian $H^{DC}$ by solving the equation
\begin{eqnarray}
(H^{DC}-E_n \mp \omega ) | \Psi_n^{(\pm)} \rangle = -D | \Psi_n\rangle.
\label{pol3}
\end{eqnarray}

\begin{table}[t]
\caption{\label{e1mat} Absolute values of the E1 matrix elements in the Li, Na, K, and Rb atoms in $ea_0$. Those are  
extracted from the measured quantities are given in bold fonts; otherwise they are calculated using the CCSD(T) method. 
Estimated uncertainties in the CCSD(T) results are given in the parentheses.}
\begin{ruledtabular}
\begin{tabular}{lclc}
Transition & E1 mat.el. & Transition & E1 mat. el. \\
\hline
Li & & Na &  \\
\hline
$2s_{1/2} \rightarrow 2p_{1/2}$  &  3.318(4)& $\bf{3s_{1/2} \rightarrow 3p_{1/2}}$   & \textbf{3.5246(23)}\\ 
$2s_{1/2} \rightarrow 3p_{1/2}$  & 0.182(2) & $3s_{1/2} \rightarrow 4p_{1/2}$   & 0.304(2)\\
$2s_{1/2} \rightarrow 4p_{1/2}$  & 0.159(2) & $3s_{1/2} \rightarrow 5p_{1/2}$   & 0.107(1)\\
$2s_{1/2} \rightarrow 5p_{1/2}$  & 0.119(4) & $3s_{1/2} \rightarrow 6p_{1/2}$   & 0.056(2)\\
$2s_{1/2} \rightarrow 6p_{1/2}$  & 0.092(2) & $3s_{1/2} \rightarrow 7p_{1/2}$   & 0.035(2)\\
$2s_{1/2} \rightarrow 7p_{1/2}$  & 0.072(1) & $3s_{1/2} \rightarrow 8p_{1/2}$   & 0.026(2)\\
$2s_{1/2} \rightarrow 2p_{3/2}$  & 4.692(5) & $\bf{3s_{1/2} \rightarrow 3p_{3/2}}$   & \textbf{4.9838(4)}\\\\
$2s_{1/2} \rightarrow 3p_{3/2}$  & 0.257(2) & $3s_{1/2} \rightarrow 4p_{3/2}$   & 0.434(2)\\
$2s_{1/2} \rightarrow 4p_{3/2}$  & 0.225(2) & $3s_{1/2} \rightarrow 5p_{3/2}$   & 0.153(2)\\
$2s_{1/2} \rightarrow 5p_{3/2}$  & 0.169(4) & $3s_{1/2} \rightarrow 6p_{3/2}$   & 0.081(2)\\
$2s_{1/2} \rightarrow 6p_{3/2}$  & 0.130(2) & $3s_{1/2} \rightarrow 7p_{3/2}$   & 0.051(2)\\
$2s_{1/2} \rightarrow 7p_{3/2}$  & 0.102(1) & $3s_{1/2} \rightarrow 8p_{3/2}$   & 0.037(2)\\
\hline
K & & Rb & \\
\hline
$\bf{4s_{1/2} \rightarrow 4p_{1/2}}$   & \textbf{4.131(20)} &  $\bf{5s_{1/2} \rightarrow 5p_{1/2}}$  & \textbf{4.227(6)}  \\ 
$4s_{1/2} \rightarrow 5p_{1/2}$   & 0.282(6)  &  $5s_{1/2} \rightarrow 6p_{1/2}$  & 0.342(2) \\
$4_{1/2} \rightarrow  6p_{1/2}$    & 0.087(5) &  $5s_{1/2} \rightarrow 7p_{1/2}$  & 0.118(1) \\
$4s_{1/2} \rightarrow 7p_{1/2}$   & 0.041(5)  &  $5s_{1/2} \rightarrow 8p_{1/2}$  & 0.061(5) \\
$4s_{1/2} \rightarrow 8p_{1/2}$   & 0.023(3)  &  $5s_{1/2} \rightarrow 9p_{1/2}$  & 0.046(3)  \\
$4s_{1/2} \rightarrow 9p_{1/2}$   & 0.016(3)  & & \\
$\bf{4s_{1/2} \rightarrow 4p_{3/2}}$   & \textbf{5.800(8)} & $\bf{5s_{1/2} \rightarrow 5p_{3/2}}$  & \textbf{5.977(9)}  \\
$4s_{1/2} \rightarrow 5p_{3/2}$   & 0.416(6)  & $5s_{1/2} \rightarrow 6p_{3/2}$  & 0.553(3) \\
$4s_{1/2} \rightarrow 6p_{3/2}$   & 0.132(6)  & $5s_{1/2} \rightarrow 7p_{3/2}$  & 0.207(2) \\
$4s_{1/2} \rightarrow 7p_{3/2}$   & 0.064(5)  & $5s_{1/2} \rightarrow 8p_{3/2}$  & 0.114(2) \\
$4s_{1/2} \rightarrow 8p_{3/2}$   & 0.038(3)  & $5s_{1/2} \rightarrow 9p_{3/2}$  & 0.074(2) \\
$4s_{1/2} \rightarrow 9p_{3/2}$   & 0.027(3)  & &\\
\end{tabular}   
\end{ruledtabular}
\end{table}
The advantage of using expression given by Eq. (\ref{pol1}) to evaluate the dynamic
polarizability is that the E1 matrix elements for many important transitions that are
predominantly contributing to the polarizabilities are now well studied and their
values are known to quite reasonable accuracy \cite{arora-sahoo1,arora-sahoo2,safronova-li,arora1,pol-andrei}. Use of these matrix
elements along with the experimental energies will certainly give more precise
contributions from these matrix elements to $\alpha$. However, the limitation 
of this sum-over-states approach is that it cannot estimate contributions from the
core electrons and can only take into account a few low-lying intermediate states.
It has been found in the previous studies that contributions with the core orbitals
and high-lying intermediate states (tail) are small compared to the low-lying intermediate 
states (e.g. see \cite{arora-sahoo1,arora-sahoo2} and {\it references therein}). Therefore, we employ a third order many-body perturbation theory (MBPT(3) method) as described in ~\cite{arora-sahoo1,arora-sahoo3}
to determine the core and tail contributions. 

Among the important E1 matrix elements between the low-lying states, the matrix elements for few primary transitions in the Na, K, and Rb atoms have been obtained using a fitting procedure from the precise measurements of the lifetimes and static dipole
polarizabilities of the first few low-lying excited states as have been given in 
\cite{arora-sahoo1,arora-sahoo2}. For instance, the E1 matrix elements of the
$3s-3p_{1/2,3/2}$ transitions are taken from the complied data list of Ref.~\cite{volz}. The other important matrix elements whose values were not deducible accurately
from the measured quantities are evaluated by employing
a relativistic coupled-cluster (RCC) theory. In our RCC method,
we express the atomic wave function with the valence electron $v$ as
\begin{eqnarray}
|\Psi_v \rangle & = & e^T \{1+S_v\} |\Phi_v \rangle ,
\label{cc2}
\end{eqnarray}
where $| \Phi_v \rangle$ is the Dirac-Fock (DF) wave function and
$T$ and $S_v$ operators account the correlation effects to all orders through the excitations of the electrons from the
core orbitals alone and from the valence orbital together from the core orbitals, respectively.
We consider here the singly and doubly excited configurations with important triple
excited configurations in the well-known CCSD(T) method framework for calculating the atomic
wave functions.

\begin{table*}[t]
\caption{\label{pol}
Static dipole polarizabilities (in a.u.) of the ground states of the Li, Na, K and Rb  alkali atoms and their comparison with the precisely
available experimental results. Values used in the single oscillator model (SOM) for the evaluation of the dynamic polarizabilities in the
previous works are also given at the bottom of the table.}
\begin{ruledtabular}
\begin{tabular}{lcccc}
Contribution & Li & Na & K & Rb\\
\hline
              $\alpha_{v}$    & 162.6    & 161.4 	 &  284.3   & 309.3	\\
 	  $\alpha_{c}$ &  0.22    &	0.9	     &	5.5		& 9.1 	\\
              $\alpha_{cv}$   & $\sim 0$ & $\sim 0$  & -0.13  	& $ -0.26$ \\
              $\alpha_{\rm{tail}}$ & 1.2      & 0.08      &  0.06    &  0.11   \\
				Total         & 164.1(7) &  162.4(2) &  289.8(6)&  318.3(6)\\
Experiment   & 164.2(11)$^a$ & 162.7(8)$^b$  & 290.58(1.42)$^c$  &  318.79(1.42)$^d$ \\
Values used in SOM		&       & 162.7(8)$^e$ & 293.6(6.1)$^f$ & 319.9(6.1)$^f$\\
\end{tabular}
\end{ruledtabular}
$^a$Ref.~\cite{li-exp}, $^b$Ref.~\cite{na-exp}, $^c$Ref.~\cite{k-exp}, $^d$Ref.~\cite{rb-exp}, $^e$Ref.~\cite{13} \\
$^f$weighted average from Ref.~\cite{14} and \cite{15}
\end{table*}
We calculate the E1 reduced matrix elements between the states $| \Psi_f \rangle$ and 
$| \Psi_i \rangle$ to be used in the sum-over-states approach using the following RCC
expression
\begin{eqnarray}
\langle \Psi_f || D || \Psi_i \rangle &=& \frac{\langle \Phi_f || \{ 1+ S_f^{\dagger}\} \overline{D } \{ 1+ S_i\} ||\Phi_i\rangle}{ \sqrt{{\cal N}_f {\cal N}_i}}, 
\end{eqnarray}
where $\overline{ D}=e^{T^{\dagger}} D e^T$ and ${\cal N}_v = \langle \Phi_v |
e^{T^{\dagger}} e^T + S_v^{\dagger} e^{T^{\dagger}} e^T S_v |\Phi_v\rangle$ 
involve two non-truncating series in the above expression. Calculation 
procedures of these expressions are discussed elsewhere in detail 
\cite{mukherjee,sahoo2}. 

In the SOM, for example used in Ref.~\cite{caride,som1}, to calculate the $C_3(a)$ coefficients, given for the evaluation of the
dynamic polarizabilities $\alpha(\omega)$ with the imaginary frequencies $\omega$ as
\begin{equation}
\alpha(\iota\omega)=\frac{\alpha(0)}{1+\frac{\omega^2}{\omega_o^2}},
\end{equation}
where $\alpha(0)$ is the static dipole polarizability (listed in table~\ref{pol}) and $\omega_0$ is the characteristic frequency of the atom. The static polarizabilities
and the characteristic frequencies are generally atom dependent.

\begin{figure}[t]
\includegraphics[scale=0.7]{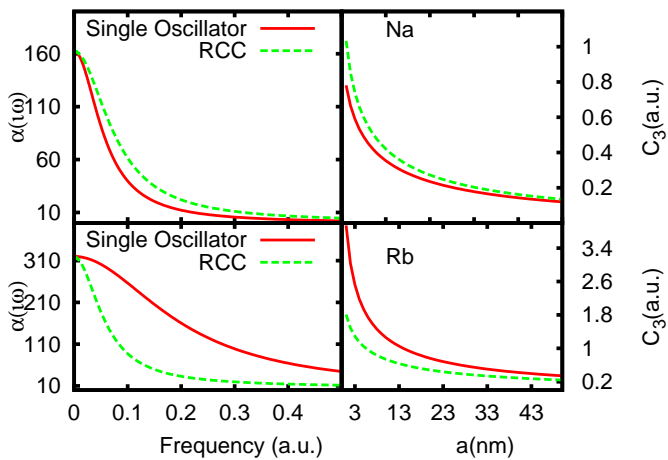}
 \caption{Dynamic polarizabilities (left half) and $C_3$ coefficients (right half) of the Na and Rb atoms interacting with a graphene layer in the Dirac model calculated using the dynamic polarizabilities obtained from the RCC calculations and from the single oscillator model (SOM).}
 \label{fig-so-rcc}     
\end{figure}

\section{Results and Discussion}\label{sec4}
As mentioned in the text above, we use the most precise values of the E1 matrix elements compiled 
in Refs.~\cite{arora-sahoo1,arora-sahoo2} for the few important $ns-np$ transitions of the Na, K, and Rb atoms. These values
are given in bold fonts in Table~\ref{e1mat}. In the same table, we also present E1 matrix elements
for other transitions calculated using our CCSD(T) method 
required for the evaluation of the polarizabilities that are already discussed in our earlier works \cite{arora-sahoo1,arora-sahoo2}. In Table \ref{pol}, we present contributions
to the static dipole polarizabilities of the considered atoms obtained using these matrix elements as valence contributions
$\alpha_v$. Contributions from the core electron excitations and correlations among the core 
electrons with the valence electron of the corresponding atoms are given as core ($\alpha_{c}$)
and core-valence ($\alpha_{cv}$), respectively, in the same table. Also, contributions from the
higher excited states whose matrix elements are not included in the determination of $\alpha_v$
are given as $\alpha_{\rm{tail}}$. All these latter three contributions are estimated using the 
MBPT(3) method in the framework as described by Eq. (\ref{pol2}). In the same table, we also give our
final polarizability results for the ground states of the alkali atoms and compare
them with the precisely available experimental results and the experimental results used in SOM~\cite{dirac-hydro2}. This
table clearly testifies the preciseness of our estimated static polarizabilities
and ensures the quality of the dynamic polarizabilities that are obtained using
these calculations. In order to compare our dynamic polarizability results with the SOM, we plot them in the left half of Fig.~\ref{fig-so-rcc}. Employing our dynamic polarizabilities
we calculate the $C_3$ coefficients for the interactions of the Na and Rb atoms with
the graphene layer. These values are plotted for the Dirac model in the right half of Fig.~\ref{fig-so-rcc}. We have also 
plotted the $C_3$ values using the dynamic polarizabilities obtained from the SOM with our calculated static 
dipole polarizabilities in the same figure. Our results for the Na atom using polarizabilities calculated using 
SOM are considerably different from the results given in Ref.~\cite{dirac-hydro}, for instance value of $C_3$(a=5 $nm$ 
is approximately equal to 1 a.u. 
from our Fig.~\ref{fig-so-rcc}, however, it is approximately 0.6 a.u. in Ref.~\cite{dirac-hydro} (see Fig. 4 of Ref.~\cite{dirac-hydro}). The discrepancy is owing to 
the fact that our dynamic polarizability values are different but more accurate from the values used by them in their calculations. 
Moreover, their predictions of the Dirac model were over estimated by a factor of 1.5 due to an error in the computer program 
(as has been clarified in~\cite{dirac-hydro2}). As can be seen, there seem to be significant differences in the 
results specially in the heavier atoms like Rb which suggest the need for more accurate polarizability results in such 
type of calculations specially for the heavier atoms.

\begin{figure*}[htpb]
 \includegraphics[width=15 cm,height=7.5 cm]{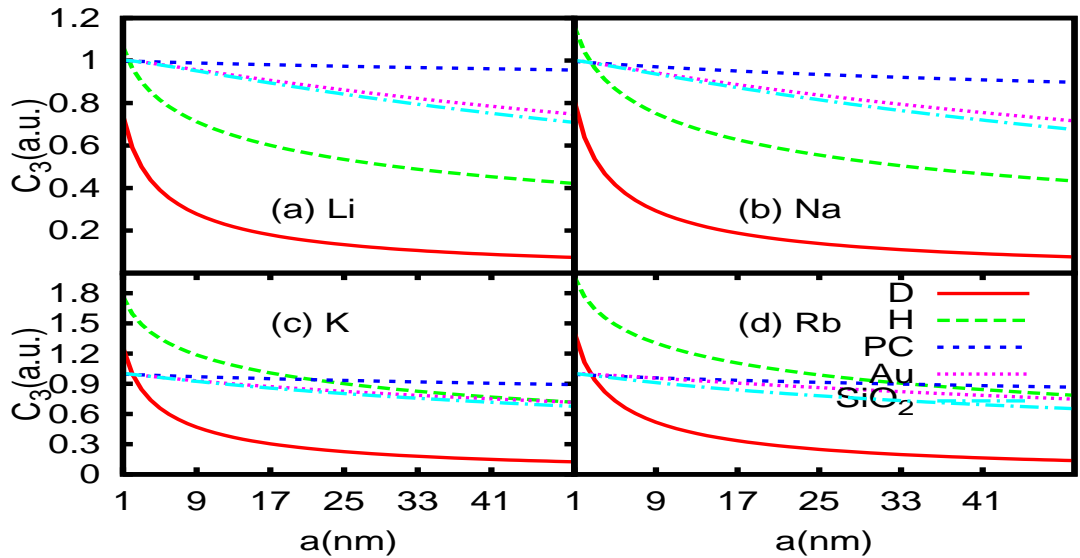}
  \caption{The $C_3$ coefficients (in a.u.) using both the Dirac (shown in solid line) and hydrodynamic (shown in long-dashed line)
 models for the alkali atoms as a function of the atom-layer separation distance $a$ interacting with the graphene 
 layer along with the results for a perfect conductor (shown in short-dashed line), Au (shown in dotted line) and SiO$_2$ (shown in dotted-dashed line).}
  \label{comp}     
\end{figure*}

\begin{figure}[h]
 \includegraphics[scale=0.7]{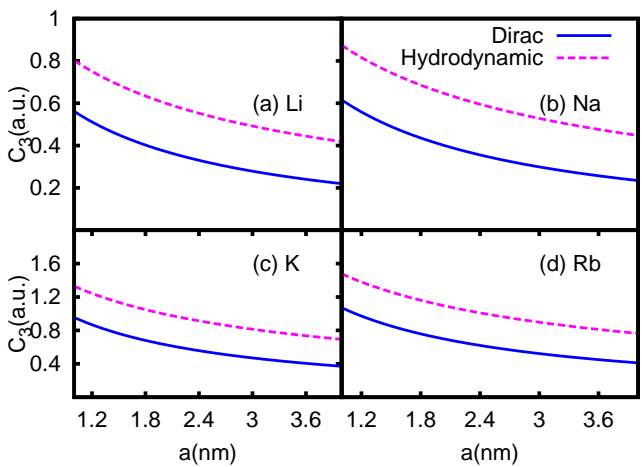}
  \caption{The $C_3$ coefficients (in a.u.) using both the Dirac (solid line) and hydrodynamic (dashed line) 
  models for the alkali atoms as a function of the atom-layer separation distance '$a$' interacting with CNT of radius $R=2 \ nm$.}
  \label{comp-cnt}     
\end{figure}

Next, we use our polarizability values to determine $C_3$ coefficients for the interaction between the
alkali atoms with the graphene layer and CNT using the hydrodynamic and Dirac model by substituting 
expressions given by Eq. (\ref{eq-h}) and Eq. (\ref{eq-d}), respectively,
in Eq. (\ref{eq-c3g}) and Eq. (\ref{eq-c3cnt}). We plot these values against different separation 
distances $a$ in Fig.~\ref{comp} for all the considered atoms using both the models and $C_3$
coefficients for a perfect conducting surface, Au and SiO$_2$ wall that were studied in our previous work \cite{bindiya3}.
From this figure, it can be observed that there are discrepancies in
the results obtained using the hydrodynamic model and Dirac model for graphene. From the physical ground
we argue that results obtained from the Dirac model are more accurate~\cite{nano1}, and the hydrodynamic
model appears to over estimate the results in graphene. As can be seen
from the figure that the interaction between an atom and graphene layer is appreciable only 
at small distances and reaches a negligible value for large separations. 
As expected the interaction of the alkali atoms is strongest with a perfect conductor for the same separation distance as compared to the interaction of the atom with a graphene layer.
However, an interesting observation can be inferred from this figure is that the $C_3$ coefficients for the K and Rb atoms interacting with the graphene layer are more 
than the interaction of atoms with the theoretically presumed ideal conductor at 
very small separation distances (say around $1-3 \ nm$). This drives us to arrive
at the conclusion that at these distances, graphene can offer tighter potentials to K and Rb atoms which can find applications in a number of experiments. 
A similar graph comparing the Dirac and hydrodynamic models for interaction between atoms and CNT with $R=2 \ nm$ as a function of separation distance '$a$' is shown in 
Fig.~\ref{comp-cnt}. The estimated $C_3$ coefficients for various atoms are shown for separation distance between 1 to 5 $nm$. For the above chosen radius of CNT and at this separation range, 
it has been observed that the exact results and the PFA results do not deviate much with respect to each other \cite{cylinder,nano1}.

\begin{table}[t]
\caption{\label{cnt}Calculated (column labeled I) and fitted (column labeled II) values obtained using 
Eq.~(\ref{fit-n}) for $C_3$ coefficients (in a.u.) for the atom-CNT interaction. Values for $R$ and a are given in $nm$.}
\begin{tabular}{ccccccccc}
\hline
 & \multicolumn{8}{c}{Li} \\
\hline
a & \multicolumn{2}{c}{R=2} &\multicolumn{2}{c}{R=4} & \multicolumn{2}{c}{R=6} & \multicolumn{2}{c}{R=8}\\
\hline
    &  I  & II   &  I    & II    &  I    & II    &  I    & II    \\
     1.0 & 0.71 & 0.681 & 0.797 & 0.775 & 0.836 & 0.833 & 0.857 & 0.855 \\
1.5 & 0.581 & 0.582 & 0.679 & 0.676 & 0.724  & 0.734 & 0.75 & 0.756 \\
2.0 & 0.493 & 0.497 & 0.592 & 0.591 & 0.64 & 0.648 & 0.668 & 0.671 \\
2.5& 0.429 & 0.427 & 0.525 & 0.52 & 0.574 & 0.578 & 0.603 & 0.6 \\
3.0 & 0.379 & 0.371 & 0.471 & 0.464 & 0.519 & 0.522 & 0.55 & 0.544 \\
3.5 & 0.339 & 0.329 & 0.426 & 0.422 & 0.474 & 0.48 & 0.505 & 0.502\\ 
4.0 & 0.306 & 0.302 & 0.389 & 0.395 & 0.436 & 0.453 & 0.466 & 0.475 \\

\hline
 & \multicolumn{8}{c}{Na} \\
\hline

 1.0 & 0.782 & 0.757 & 0.884 & 0.858 & 0.927 & 0.921 & 0.95 & 0.945 \\ 
1.5 & 0.638 & 0.642 & 0.747 & 0.744& 0.796 & 0.806 & 0.824 & 0.831 \\ 
 2.0 & 0.539 & 0.544 & 0.647 & 0.646 & 0.699 & 0.709 & 0.73 & 0.733 \\ 
 2.5 & 0.466 & 0.47 & 0.571 & 0.565 & 0.624 & 0.628 & 0.656 & 0.652\\
 3.0 & 0.41 & 0.4 & 0.51 & 0.502 & 0.563 & 0.565 & 0.596 & 0.589\\
3.5 & 0.366 & 0.354 & 0.461 & 0.456 & 0.512 & 0.519 & 0.545 & 0.543 \\
 4.0 & 0.33 & 0.325 & 0.42 & 0.427 & 0.47 & 0.49 & 0.502 & 0.514\\
\hline
 & \multicolumn{8}{c}{K} \\
\hline

1.0 & 1.212 & 1.172 & 1.37 & 1.33 & 1.437 & 1.428 & 1.473 & 1.465  \\ 
1.5 & 0.99 & 0.995 &1.16 & 1.154 & 1.235  & 1.252 & 1.278 & 1.289 \\
2.0 & 0.838 & 0.845 & 1.0 & 1.0 & 1.087 &1.1 & 1.135 & 1.139  \\
2.5& 0.726 & 0.722 & 0.889 & 0.88 & 0.971 & 0.978 & 1.022 & 1.015\\
3.0 & 0.64 & 0.675 & 0.796 & 0.783 & 0.878 & 0.881 & 0.929 & 0.919 \\
3.5 & 0.572 & 0.554 & 0.72 & 0.712 & 0.801 & 0.811 & 0.852 & 0.848 \\
4.0 & 0.516 & 0.51 & 0.657 & 0.669 & 0.736 & 0.767 & 0.787 & 0.804\\

\hline
 & \multicolumn{8}{c}{Rb} \\
\hline

1.0 & 1.371 & 1.325 & 1.548 & 1.502 & 1.624 & 1.611 & 1.665 & 1.653  \\ 
1.5 & 1.113 & 1.121 &1.302 & 1.298 & 1.388  & 1.407 & 1.438 & 1.449 \\
2.0 & 0.939 & 0.948 & 1.127& 1.125 & 1.217 &1.235 & 1.271 & 1.277 \\
2.5& 0.8 & 0.807 & 0.993 & 0.984 & 1.04 & 1.093 & 1.097 & 1.135\\
3.0 & 0.714 & 0.696 & 0.888 & 0.873 & 0.979 & 0.983 & 1.037 & 1.025 \\
3.5 & 0.637 & 0.616 & 0.802 & 0.794 & 0.892 & 0.903 & 0.949 & 0.945 \\
4.0 & 0.574 & 0.568 & 0.731 & 0.745 & 0.818 & 0.855 & 0.875 & 0.897\\
\hline
\end{tabular} 

\end{table}

\begin{table}[t]
\caption{\label{fit} Fitting parameters for $C_3$(a) coefficients with a graphene layer and CNT.}
\begin{ruledtabular}
\begin{tabular}{ccccc}
 {Graphene layer}  & Li & Na & K & Rb\\
 \hline
 A$_0$(a.u.) & 7.4355 & 7.61362 & 12.5622 & 13.7257\\
 B$_0$($nm$) & 8.36468 & 7.74636 & 8.41002 & 8.19064\\
 \hline 
{CNT} & Li & Na & K & Rb\\
\hline
C$_0$(a.u.) & 0.79556 & 0.89764 & 1.3852 & 1.5806 \\
A(a.u./$nm$) & -0.27172 & -0.31559 &  -0.48514 & -0.5628 \\
B(a.u./$nm$) & 0.07338 & 0.07988 & 0.12444 & 0.13918  \\
C(a.u./$nm^2$) & 0.02902 & 0.03436 & 0.05293 & 0.06213 \\
D(a.u./$nm^2$) & -0.00445 & -0.00484  & -0.00754 &   -0.00844\\
\end{tabular} 
\end{ruledtabular}
\end{table}
One of our motivations to carry out this study is also to find out the dependence of the atom-wall
interactions on the radius of CNT. For this purpose,
we present the results computed for the $C_3$ coefficients as a function of distance '$a$' and radius '$R$' of CNT in 
Table.~\ref{cnt}. The range for $R$ and $a$ have been chosen in accordance with the validity range of PFA. From the table, we notice 
that the $C_3$ coefficients increase slowly with the increase in the CNT radii, however the 
rate of increase is not very magnificent. With a three-fold increase in the radius, it raises the
$C_3$ coefficients only about one and half times. As expected these coefficients get stronger as 
the size of the atom increases; i.e. from Li to Rb for a given separation distance '$a$'. 
We were unable to find out any previous work to compare our results with; however, we have exercised 
the cross-checking between our results for the H atom and H$_2$ molecule independently with the results 
reported in \cite{blagov} for CNT to ascertain our calculation procedure.

A lot of research work is devoted to the experimental investigation of the behavior of the
interactions between the trapped atoms with the graphene layers or CNTs ~\cite{trapping,trapping1,trapping2,trapping3,trapping4,trapping5}. For simplification 
of reproducing the surface interaction potentials from our reported $C_3$ coefficients and
for any comparison of our results with theoretical values, we give a logistic fit for the interaction potential
of the atom-graphene layer interaction using the following form 
\begin{equation}
U(a)=\frac{A_0}{a^3(a+B_0)} , \label{fit-g}
\end{equation} 
where $A_0$ (in a.u.) and $B_0$ (in $nm$) are the fitting parameters that depend on the properties of the atom. A list 
of these fitting parameters for the Li, Na, K and Rb atoms are given in Table~\ref{fit}.

The above equation is a useful tool to predict the interaction between the alkali atoms and a graphene 
layer for any given separation distance '$a$'. 
We have used the mass gap parameter $\Delta$ value as 0.1 eV in our calculations. It
has been observed that a change in $\Delta$ value from 0.1 eV to $10^{-5}$ eV causes a change of 13\% in the 
fitting parameters. Our fitting parameters for interaction between graphene and Na atoms are considerably different 
from those calculated in Ref~\cite{dirac-hydro} ($A_0$=7.11 a.u. and $B_0$=9.77 $nm$). As mentioned previously our results are more reliable keeping in mind the error in the code in Ref.~\cite{dirac-hydro} and use of our fitting parameters is recommended in extrapolating the interaction potential for graphene-alkali atom interaction. Similarly, we also fit the $U(a,R)$ results for the interaction of
these atoms with CNT. However, a logistic equation didn't serve as a suitable fit for CNT, instead 
we use a rational Taylor equation to fit the results in the following functional form
\begin{equation}
 U(a,R)=\frac{C_0+Aa+BR+Ca^2+DR^2}{a^3}\label{fit-n}
\end{equation}
and present the respective fitting coefficients with units in Table \ref{fit} with best goodness. 
In Table~\ref{cnt}, we compare our fitted $C_3$ coefficient values (column labeled II) with those calculated 
using Eq.~(\ref{eq-c3cnt}) and Eq.~(\ref{eq-d}) (column labeled I). We see a deviation of less then 4\% at all 
the separation distances.

\section{Summary}
To summarize, we have investigated the dispersion coefficients for the atom-graphene and atom-carbon nanotube 
interactions for the Li, Na, K, and Rb atoms in this work and compared our results with the previously reported results
and against the results for the interaction of atoms with a perfect conductor. 
The interaction potentials of the alkali atoms are 
studied using both the hydrodynamic and Dirac models and their dependence on
the distance between the atom and the nanotube or graphene layer and radius of the nanotube 
are investigated. The importance of using high precision dynamic polarizability values for such calculations specially for the heavier atoms is highlighted. 
 Readily usable functional forms for the interaction potentials are suggested for the easy extrapolation and  comparison of the 
experimental results with the theoretical values. 

\section*{Acknowledgement}
The work of B.A. is supported by CSIR grant no. 03(1268)/13/EMR-II, India. H.K. acknowledges the financial support from CSIR. Computations were carried 
out using 3TFLOP HPC Cluster at Physical Research Laboratory, Ahmedabad.


\begin{thebibliography}{85}
\expandafter\ifx\csname natexlab\endcsname\relax\def\natexlab#1{#1}\fi
\expandafter\ifx\csname bibnamefont\endcsname\relax
  \def\bibnamefont#1{#1}\fi
\expandafter\ifx\csname bibfnamefont\endcsname\relax
  \def\bibfnamefont#1{#1}\fi
\expandafter\ifx\csname citenamefont\endcsname\relax
  \def\citenamefont#1{#1}\fi
\expandafter\ifx\csname url\endcsname\relax
  \def\url#1{\texttt{#1}}\fi
\expandafter\ifx\csname urlprefix\endcsname\relax\def\urlprefix{URL }\fi
\providecommand{\bibinfo}[2]{#2}
\providecommand{\eprint}[2][]{\url{#2}}

\bibitem[{\citenamefont{Agranovich and Ginzburg}(1984)}]{agra}
\bibinfo{author}{\bibfnamefont{V.~M.} \bibnamefont{Agranovich}}
  \bibnamefont{and} \bibinfo{author}{\bibfnamefont{V.~L.}
  \bibnamefont{Ginzburg}}, \emph{\bibinfo{title}{Crystal Optics with Spatial
  Dispersion}} (\bibinfo{publisher}{Springer, Berlin}, \bibinfo{year}{1984}).

\bibitem[{\citenamefont{Fermani et~al.}(2007)\citenamefont{Fermani, Scheel, and
  Knight}}]{fermani}
\bibinfo{author}{\bibfnamefont{R.}~\bibnamefont{Fermani}},
  \bibinfo{author}{\bibfnamefont{S.}~\bibnamefont{Scheel}}, \bibnamefont{and}
  \bibinfo{author}{\bibfnamefont{P.~L.} \bibnamefont{Knight}},
  \bibinfo{journal}{Phys. Rev. A} \textbf{\bibinfo{volume}{75}},
  \bibinfo{pages}{062905} (\bibinfo{year}{2007}).

\bibitem[{\citenamefont{Friedrich et~al.}(2002)\citenamefont{Friedrich, Jacoby,
  and Meister}}]{friedrich}
\bibinfo{author}{\bibfnamefont{H.}~\bibnamefont{Friedrich}},
  \bibinfo{author}{\bibfnamefont{G.}~\bibnamefont{Jacoby}}, \bibnamefont{and}
  \bibinfo{author}{\bibfnamefont{C.~G.} \bibnamefont{Meister}},
  \bibinfo{journal}{Phys. Rev. A} \textbf{\bibinfo{volume}{65}},
  \bibinfo{pages}{032902} (\bibinfo{year}{2002}).

\bibitem[{\citenamefont{Novoselov et~al.}(2005)\citenamefont{Novoselov, Geim,
  Morozov, Jiang, Katsnelson, Grigorieva, Dubonos, and Firsov}}]{intro1}
\bibinfo{author}{\bibfnamefont{K.~S.} \bibnamefont{Novoselov}},
  \bibinfo{author}{\bibfnamefont{A.~K.} \bibnamefont{Geim}},
  \bibinfo{author}{\bibfnamefont{S.~V.} \bibnamefont{Morozov}},
  \bibinfo{author}{\bibfnamefont{D.}~\bibnamefont{Jiang}},
  \bibinfo{author}{\bibfnamefont{M.~I.} \bibnamefont{Katsnelson}},
  \bibinfo{author}{\bibfnamefont{I.~V.} \bibnamefont{Grigorieva}},
  \bibinfo{author}{\bibfnamefont{S.~V.} \bibnamefont{Dubonos}},
  \bibnamefont{and} \bibinfo{author}{\bibfnamefont{A.~A.}
  \bibnamefont{Firsov}}, \bibinfo{journal}{Nature}
  \textbf{\bibinfo{volume}{438}}, \bibinfo{pages}{197} (\bibinfo{year}{2005}).

\bibitem[{\citenamefont{Novoselov et~al.}(2004)\citenamefont{Novoselov, Geim1,
  Morozov, Jiang, Zhang, Dubonos, Grigorieva, and Firsov}}]{intro2}
\bibinfo{author}{\bibfnamefont{K.~S.} \bibnamefont{Novoselov}},
  \bibinfo{author}{\bibfnamefont{A.~K.} \bibnamefont{Geim1}},
  \bibinfo{author}{\bibfnamefont{S.~V.} \bibnamefont{Morozov}},
  \bibinfo{author}{\bibfnamefont{D.}~\bibnamefont{Jiang}},
  \bibinfo{author}{\bibfnamefont{Y.}~\bibnamefont{Zhang}},
  \bibinfo{author}{\bibfnamefont{S.~V.} \bibnamefont{Dubonos}},
  \bibinfo{author}{\bibfnamefont{I.~V.} \bibnamefont{Grigorieva}},
  \bibnamefont{and} \bibinfo{author}{\bibfnamefont{A.~A.}
  \bibnamefont{Firsov}}, \bibinfo{journal}{Science}
  \textbf{\bibinfo{volume}{306}}, \bibinfo{pages}{666} (\bibinfo{year}{2004}).

\bibitem[{\citenamefont{Burress et~al.}(2010)\citenamefont{Burress, Gadipelli,
  Ford, Simmons, Zhou1, and Yildirim1}}]{intro3}
\bibinfo{author}{\bibfnamefont{J.~W.} \bibnamefont{Burress}},
  \bibinfo{author}{\bibfnamefont{S.}~\bibnamefont{Gadipelli}},
  \bibinfo{author}{\bibfnamefont{J.}~\bibnamefont{Ford}},
  \bibinfo{author}{\bibfnamefont{J.~M.} \bibnamefont{Simmons}},
  \bibinfo{author}{\bibfnamefont{W.}~\bibnamefont{Zhou1}}, \bibnamefont{and}
  \bibinfo{author}{\bibfnamefont{T.}~\bibnamefont{Yildirim1}},
  \bibinfo{journal}{Angew. Chem. Int.} \textbf{\bibinfo{volume}{49}},
  \bibinfo{pages}{8902} (\bibinfo{year}{2010}).

\bibitem[{\citenamefont{Tozzini and Pellegrini}(2013)}]{intro4}
\bibinfo{author}{\bibfnamefont{V.}~\bibnamefont{Tozzini}} \bibnamefont{and}
  \bibinfo{author}{\bibfnamefont{V.}~\bibnamefont{Pellegrini}},
  \bibinfo{journal}{Phys. Chem. Chem. Phys.} \textbf{\bibinfo{volume}{15}},
  \bibinfo{pages}{80} (\bibinfo{year}{2013}).

\bibitem[{\citenamefont{Spyrou et~al.}(2013)\citenamefont{Spyrou, Gournis, and
  Rudolf}}]{intro5}
\bibinfo{author}{\bibfnamefont{K.}~\bibnamefont{Spyrou}},
  \bibinfo{author}{\bibfnamefont{D.}~\bibnamefont{Gournis}}, \bibnamefont{and}
  \bibinfo{author}{\bibfnamefont{P.}~\bibnamefont{Rudolf}},
  \bibinfo{journal}{ECS J. Solid State Sci. Technol.}
  \textbf{\bibinfo{volume}{2}}, \bibinfo{pages}{M3160} (\bibinfo{year}{2013}).

\bibitem[{\citenamefont{Schedin et~al.}(2007)\citenamefont{Schedin, Geim,
  Morozov, Hill, Blake, Katsnelson, and Novoselov}}]{intro6}
\bibinfo{author}{\bibfnamefont{F.}~\bibnamefont{Schedin}},
  \bibinfo{author}{\bibfnamefont{A.~K.} \bibnamefont{Geim}},
  \bibinfo{author}{\bibfnamefont{S.~V.} \bibnamefont{Morozov}},
  \bibinfo{author}{\bibfnamefont{E.~W.} \bibnamefont{Hill}},
  \bibinfo{author}{\bibfnamefont{P.}~\bibnamefont{Blake}},
  \bibinfo{author}{\bibfnamefont{M.~I.} \bibnamefont{Katsnelson}},
  \bibnamefont{and} \bibinfo{author}{\bibfnamefont{K.~S.}
  \bibnamefont{Novoselov}}, \bibinfo{journal}{Nature Materials}
  \textbf{\bibinfo{volume}{6}}, \bibinfo{pages}{652} (\bibinfo{year}{2007}).

\bibitem[{\citenamefont{Zhang et~al.}(2013)\citenamefont{Zhang, Liu, Meng, Li,
  Liang, Hu, and Wang}}]{intro7}
\bibinfo{author}{\bibfnamefont{B.~Y.} \bibnamefont{Zhang}},
  \bibinfo{author}{\bibfnamefont{T.}~\bibnamefont{Liu}},
  \bibinfo{author}{\bibfnamefont{B.}~\bibnamefont{Meng}},
  \bibinfo{author}{\bibfnamefont{X.}~\bibnamefont{Li}},
  \bibinfo{author}{\bibfnamefont{G.}~\bibnamefont{Liang}},
  \bibinfo{author}{\bibfnamefont{X.}~\bibnamefont{Hu}}, \bibnamefont{and}
  \bibinfo{author}{\bibfnamefont{Q.~J.} \bibnamefont{Wang}},
  \bibinfo{journal}{Nature Commincations} \textbf{\bibinfo{volume}{4}},
  \bibinfo{pages}{1811} (\bibinfo{year}{2013}).

\bibitem[{\citenamefont{Shen et~al.}(2012)\citenamefont{Shen, Zhang, Liu, and
  Zhang}}]{intro8}
\bibinfo{author}{\bibfnamefont{H.}~\bibnamefont{Shen}},
  \bibinfo{author}{\bibfnamefont{L.}~\bibnamefont{Zhang}},
  \bibinfo{author}{\bibfnamefont{M.}~\bibnamefont{Liu}}, \bibnamefont{and}
  \bibinfo{author}{\bibfnamefont{Z.}~\bibnamefont{Zhang}},
  \bibinfo{journal}{Theranostics} \textbf{\bibinfo{volume}{2}},
  \bibinfo{pages}{283} (\bibinfo{year}{2012}).

\bibitem[{\citenamefont{Nguyen and Berry}(2012)}]{intro9}
\bibinfo{author}{\bibfnamefont{P.}~\bibnamefont{Nguyen}} \bibnamefont{and}
  \bibinfo{author}{\bibfnamefont{V.}~\bibnamefont{Berry}}, \bibinfo{journal}{J.
  Phys. Chem. Lett.} \textbf{\bibinfo{volume}{3}}, \bibinfo{pages}{1024}
  (\bibinfo{year}{2012}).

\bibitem[{\citenamefont{Churkin et~al.}(2010)\citenamefont{Churkin, Fedortsov,
  Klimchitskaya, and Yurova}}]{dirac-hydro}
\bibinfo{author}{\bibfnamefont{Y.~V.} \bibnamefont{Churkin}},
  \bibinfo{author}{\bibfnamefont{A.~B.} \bibnamefont{Fedortsov}},
  \bibinfo{author}{\bibfnamefont{G.~L.} \bibnamefont{Klimchitskaya}},
  \bibnamefont{and} \bibinfo{author}{\bibfnamefont{V.~A.}
  \bibnamefont{Yurova}}, \bibinfo{journal}{Phys. Rev. B}
  \textbf{\bibinfo{volume}{82}}, \bibinfo{pages}{165433}
  (\bibinfo{year}{2010}).

\bibitem[{\citenamefont{Shimizu}(2001)}]{qr1}
\bibinfo{author}{\bibfnamefont{F.}~\bibnamefont{Shimizu}},
  \bibinfo{journal}{Phys. Rev. Lett.} \textbf{\bibinfo{volume}{86}},
  \bibinfo{pages}{987} (\bibinfo{year}{2001}).

\bibitem[{\citenamefont{Druzhinina and DeKieviet}(2003)}]{qr2}
\bibinfo{author}{\bibfnamefont{V.}~\bibnamefont{Druzhinina}} \bibnamefont{and}
  \bibinfo{author}{\bibfnamefont{M.}~\bibnamefont{DeKieviet}},
  \bibinfo{journal}{Phys. Rev. Lett.} \textbf{\bibinfo{volume}{91}},
  \bibinfo{pages}{193202} (\bibinfo{year}{2003}).

\bibitem[{\citenamefont{Lin et~al.}(2004)\citenamefont{Lin, Teper, Chin, and
  Vuletic}}]{qr3}
\bibinfo{author}{\bibfnamefont{Y.~J.} \bibnamefont{Lin}},
  \bibinfo{author}{\bibfnamefont{I.}~\bibnamefont{Teper}},
  \bibinfo{author}{\bibfnamefont{C.}~\bibnamefont{Chin}}, \bibnamefont{and}
  \bibinfo{author}{\bibfnamefont{V.}~\bibnamefont{Vuletic}},
  \bibinfo{journal}{Phys. Rev. Lett.} \textbf{\bibinfo{volume}{92}},
  \bibinfo{pages}{050404} (\bibinfo{year}{2004}).

\bibitem[{\citenamefont{Kaur et~al.}(2013)\citenamefont{Kaur, Gupta, and
  Dharamvir}}]{intro10}
\bibinfo{author}{\bibfnamefont{G.}~\bibnamefont{Kaur}},
  \bibinfo{author}{\bibfnamefont{S.}~\bibnamefont{Gupta}}, \bibnamefont{and}
  \bibinfo{author}{\bibfnamefont{K.}~\bibnamefont{Dharamvir}},
  \bibinfo{journal}{International Journal of Physics and Research}
  \textbf{\bibinfo{volume}{3}}, \bibinfo{pages}{1} (\bibinfo{year}{2013}).

\bibitem[{\citenamefont{Lee et~al.}(1997)\citenamefont{Lee, Kim, Fischer,
  Thess, and Smalley}}]{intro12}
\bibinfo{author}{\bibfnamefont{R.~S.} \bibnamefont{Lee}},
  \bibinfo{author}{\bibfnamefont{H.~J.} \bibnamefont{Kim}},
  \bibinfo{author}{\bibfnamefont{J.~E.} \bibnamefont{Fischer}},
  \bibinfo{author}{\bibfnamefont{A.}~\bibnamefont{Thess}}, \bibnamefont{and}
  \bibinfo{author}{\bibfnamefont{R.~E.} \bibnamefont{Smalley}},
  \bibinfo{journal}{Nature} \textbf{\bibinfo{volume}{388}},
  \bibinfo{pages}{255} (\bibinfo{year}{1997}).

\bibitem[{\citenamefont{Lee et~al.}(2000)\citenamefont{Lee, Kim, Fischer,
  Lefebvre, Radosavljevic, Hone, and Johnson}}]{intro13}
\bibinfo{author}{\bibfnamefont{R.~S.} \bibnamefont{Lee}},
  \bibinfo{author}{\bibfnamefont{H.~J.} \bibnamefont{Kim}},
  \bibinfo{author}{\bibfnamefont{J.~E.} \bibnamefont{Fischer}},
  \bibinfo{author}{\bibfnamefont{J.}~\bibnamefont{Lefebvre}},
  \bibinfo{author}{\bibfnamefont{M.}~\bibnamefont{Radosavljevic}},
  \bibinfo{author}{\bibfnamefont{J.}~\bibnamefont{Hone}}, \bibnamefont{and}
  \bibinfo{author}{\bibfnamefont{A.~T.} \bibnamefont{Johnson}},
  \bibinfo{journal}{Phys. Rev. B} \textbf{\bibinfo{volume}{61}},
  \bibinfo{pages}{4526} (\bibinfo{year}{2000}).

\bibitem[{\citenamefont{Gao et~al.}(1998)\citenamefont{Gao, Cagin, and
  Goddard}}]{intro14}
\bibinfo{author}{\bibfnamefont{G.}~\bibnamefont{Gao}},
  \bibinfo{author}{\bibfnamefont{T.}~\bibnamefont{Cagin}}, \bibnamefont{and}
  \bibinfo{author}{\bibfnamefont{W.~A.} \bibnamefont{Goddard}},
  \bibinfo{journal}{Phys. Rev. Lett.} \textbf{\bibinfo{volume}{80}},
  \bibinfo{pages}{5556} (\bibinfo{year}{1998}).

\bibitem[{\citenamefont{Ataca et~al.}(2008)\citenamefont{Ataca, Aktürk,
  Ciraci, and Ustunel}}]{ataca}
\bibinfo{author}{\bibfnamefont{C.}~\bibnamefont{Ataca}},
  \bibinfo{author}{\bibfnamefont{E.}~\bibnamefont{Aktürk}},
  \bibinfo{author}{\bibfnamefont{S.}~\bibnamefont{Ciraci}}, \bibnamefont{and}
  \bibinfo{author}{\bibfnamefont{H.}~\bibnamefont{Ustunel}},
  \bibinfo{journal}{Appl. Phys. Lett.} \textbf{\bibinfo{volume}{93}},
  \bibinfo{pages}{043123} (\bibinfo{year}{2008}).

\bibitem[{\citenamefont{Du et~al.}(2010)\citenamefont{Du, Zhu, and Smith}}]{du}
\bibinfo{author}{\bibfnamefont{A.}~\bibnamefont{Du}},
  \bibinfo{author}{\bibfnamefont{Z.}~\bibnamefont{Zhu}}, \bibnamefont{and}
  \bibinfo{author}{\bibfnamefont{S.~C.} \bibnamefont{Smith}},
  \bibinfo{journal}{J. Am. Chem. Soc.} \textbf{\bibinfo{volume}{132}},
  \bibinfo{pages}{2876} (\bibinfo{year}{2010}).

\bibitem[{\citenamefont{kai Konga and wang Chen}(2013)}]{konga}
\bibinfo{author}{\bibfnamefont{X.}~\bibnamefont{kai Konga}} \bibnamefont{and}
  \bibinfo{author}{\bibfnamefont{Q.}~\bibnamefont{wang Chen}},
  \bibinfo{journal}{Phys. Chem. Chem. Phys.} \textbf{\bibinfo{volume}{15}},
  \bibinfo{pages}{12982} (\bibinfo{year}{2013}).

\bibitem[{\citenamefont{Lin et~al.}(2013)\citenamefont{Lin, Peng, Xiang, Ruan,
  Yan, Natelson, and Tour}}]{jian}
\bibinfo{author}{\bibfnamefont{J.}~\bibnamefont{Lin}},
  \bibinfo{author}{\bibfnamefont{Z.}~\bibnamefont{Peng}},
  \bibinfo{author}{\bibfnamefont{C.}~\bibnamefont{Xiang}},
  \bibinfo{author}{\bibfnamefont{G.}~\bibnamefont{Ruan}},
  \bibinfo{author}{\bibfnamefont{Z.}~\bibnamefont{Yan}},
  \bibinfo{author}{\bibfnamefont{D.}~\bibnamefont{Natelson}}, \bibnamefont{and}
  \bibinfo{author}{\bibfnamefont{J.~M.} \bibnamefont{Tour}},
  \bibinfo{journal}{Phys. Chem. Chem. Phys.} \textbf{\bibinfo{volume}{15}},
  \bibinfo{pages}{12982} (\bibinfo{year}{2013}).

\bibitem[{\citenamefont{Neto et~al.}(2009)\citenamefont{Neto, Guinea, Peres,
  Novoselov, and Geim}}]{castro}
\bibinfo{author}{\bibfnamefont{A.~H.~C.} \bibnamefont{Neto}},
  \bibinfo{author}{\bibfnamefont{F.}~\bibnamefont{Guinea}},
  \bibinfo{author}{\bibfnamefont{N.~M.~R.} \bibnamefont{Peres}},
  \bibinfo{author}{\bibfnamefont{K.~S.} \bibnamefont{Novoselov}},
  \bibnamefont{and} \bibinfo{author}{\bibfnamefont{A.~K.} \bibnamefont{Geim}},
  \bibinfo{journal}{Rev. Mod. Phys.} \textbf{\bibinfo{volume}{81}},
  \bibinfo{pages}{109} (\bibinfo{year}{2009}).

\bibitem[{\citenamefont{Geim}(2009)}]{intro16}
\bibinfo{author}{\bibfnamefont{A.~K.} \bibnamefont{Geim}},
  \bibinfo{journal}{Science} \textbf{\bibinfo{volume}{324}},
  \bibinfo{pages}{1530} (\bibinfo{year}{2009}).

\bibitem[{\citenamefont{Oh et~al.}(2010)\citenamefont{Oh, Shin, and
  Ahn}}]{intro17}
\bibinfo{author}{\bibfnamefont{D.-H.} \bibnamefont{Oh}},
  \bibinfo{author}{\bibfnamefont{B.~G.} \bibnamefont{Shin}}, \bibnamefont{and}
  \bibinfo{author}{\bibfnamefont{J.~R.} \bibnamefont{Ahn}},
  \bibinfo{journal}{App. Phys. Lett.} \textbf{\bibinfo{volume}{96}},
  \bibinfo{pages}{231916} (\bibinfo{year}{2010}).

\bibitem[{\citenamefont{Ohta et~al.}(2006)\citenamefont{Ohta, Bostwick,
  Seyller, Horn, and Rotenberg}}]{intro18}
\bibinfo{author}{\bibfnamefont{T.}~\bibnamefont{Ohta}},
  \bibinfo{author}{\bibfnamefont{A.}~\bibnamefont{Bostwick}},
  \bibinfo{author}{\bibfnamefont{T.}~\bibnamefont{Seyller}},
  \bibinfo{author}{\bibfnamefont{K.}~\bibnamefont{Horn}}, \bibnamefont{and}
  \bibinfo{author}{\bibfnamefont{E.}~\bibnamefont{Rotenberg}},
  \bibinfo{journal}{Science} \textbf{\bibinfo{volume}{313}},
  \bibinfo{pages}{951} (\bibinfo{year}{2006}).

\bibitem[{\citenamefont{Iijima}(1991)}]{cnt-1}
\bibinfo{author}{\bibfnamefont{S.}~\bibnamefont{Iijima}},
  \bibinfo{journal}{Nature} \textbf{\bibinfo{volume}{354}}, \bibinfo{pages}{56}
  (\bibinfo{year}{1991}).

\bibitem[{\citenamefont{Ebbesen and Ajayan}(1992)}]{cnt-2}
\bibinfo{author}{\bibfnamefont{T.~W.} \bibnamefont{Ebbesen}} \bibnamefont{and}
  \bibinfo{author}{\bibfnamefont{P.~M.} \bibnamefont{Ajayan}},
  \bibinfo{journal}{Nature} \textbf{\bibinfo{volume}{358}},
  \bibinfo{pages}{220} (\bibinfo{year}{1992}).

\bibitem[{\citenamefont{Guo et~al.}(1995)\citenamefont{Guo, Nikolaev, Rinzler,
  Tomanek, Colbert, and Smalley}}]{cnt-3}
\bibinfo{author}{\bibfnamefont{T.}~\bibnamefont{Guo}},
  \bibinfo{author}{\bibfnamefont{P.}~\bibnamefont{Nikolaev}},
  \bibinfo{author}{\bibfnamefont{A.~G.} \bibnamefont{Rinzler}},
  \bibinfo{author}{\bibfnamefont{D.}~\bibnamefont{Tomanek}},
  \bibinfo{author}{\bibfnamefont{D.~T.} \bibnamefont{Colbert}},
  \bibnamefont{and} \bibinfo{author}{\bibfnamefont{R.~E.}
  \bibnamefont{Smalley}}, \bibinfo{journal}{J. Phys. Chem.}
  \textbf{\bibinfo{volume}{99}}, \bibinfo{pages}{10694–10697}
  (\bibinfo{year}{1995}).

\bibitem[{\citenamefont{Barberio et~al.}(2012)\citenamefont{Barberio, Barone,
  Bonanno, and Oliva}}]{cnt-pur}
\bibinfo{author}{\bibfnamefont{M.}~\bibnamefont{Barberio}},
  \bibinfo{author}{\bibfnamefont{P.}~\bibnamefont{Barone}},
  \bibinfo{author}{\bibfnamefont{A.}~\bibnamefont{Bonanno}}, \bibnamefont{and}
  \bibinfo{author}{\bibfnamefont{A.}~\bibnamefont{Oliva}}, \bibinfo{journal}{J.
  Nanosci. Nanotechnol.} \textbf{\bibinfo{volume}{12}}, \bibinfo{pages}{5039}
  (\bibinfo{year}{2012}).

\bibitem[{\citenamefont{Radosavljevic et~al.}(2004)\citenamefont{Radosavljevic,
  Appenzeller, Avouris, and Knoch}}]{cnt-fet}
\bibinfo{author}{\bibfnamefont{M.}~\bibnamefont{Radosavljevic}},
  \bibinfo{author}{\bibfnamefont{J.}~\bibnamefont{Appenzeller}},
  \bibinfo{author}{\bibfnamefont{P.}~\bibnamefont{Avouris}}, \bibnamefont{and}
  \bibinfo{author}{\bibfnamefont{J.}~\bibnamefont{Knoch}},
  \bibinfo{journal}{Appl. Phys. Lett.} \textbf{\bibinfo{volume}{84}},
  \bibinfo{pages}{3693} (\bibinfo{year}{2004}).

\bibitem[{\citenamefont{Bogicevic et~al.}(2000)\citenamefont{Bogicevic,
  Ovesson, Hyldgaard, Lundqvist, Brune, and Jennison}}]{dft1}
\bibinfo{author}{\bibfnamefont{A.}~\bibnamefont{Bogicevic}},
  \bibinfo{author}{\bibfnamefont{S.}~\bibnamefont{Ovesson}},
  \bibinfo{author}{\bibfnamefont{P.}~\bibnamefont{Hyldgaard}},
  \bibinfo{author}{\bibfnamefont{B.~I.} \bibnamefont{Lundqvist}},
  \bibinfo{author}{\bibfnamefont{H.}~\bibnamefont{Brune}}, \bibnamefont{and}
  \bibinfo{author}{\bibfnamefont{D.~R.} \bibnamefont{Jennison}},
  \bibinfo{journal}{Phys. Rev. Lett.} \textbf{\bibinfo{volume}{85}},
  \bibinfo{pages}{1910} (\bibinfo{year}{2000}).

\bibitem[{\citenamefont{Jung et~al.}(2004)\citenamefont{Jung, Garcia-Gonzalez,
  Dobson, and Godby}}]{dft2}
\bibinfo{author}{\bibfnamefont{J.}~\bibnamefont{Jung}},
  \bibinfo{author}{\bibfnamefont{P.}~\bibnamefont{Garcia-Gonzalez}},
  \bibinfo{author}{\bibfnamefont{J.~F.} \bibnamefont{Dobson}},
  \bibnamefont{and} \bibinfo{author}{\bibfnamefont{R.~W.} \bibnamefont{Godby}},
  \bibinfo{journal}{Phys. Rev. B} \textbf{\bibinfo{volume}{70}},
  \bibinfo{pages}{205107} (\bibinfo{year}{2004}).

\bibitem[{\citenamefont{Dobson et~al.}(2006)\citenamefont{Dobson, White, and
  Rubio}}]{dft3}
\bibinfo{author}{\bibfnamefont{J.~F.} \bibnamefont{Dobson}},
  \bibinfo{author}{\bibfnamefont{A.}~\bibnamefont{White}}, \bibnamefont{and}
  \bibinfo{author}{\bibfnamefont{A.}~\bibnamefont{Rubio}},
  \bibinfo{journal}{Phys. Rev. Lett.} \textbf{\bibinfo{volume}{96}},
  \bibinfo{pages}{073201} (\bibinfo{year}{2006}).

\bibitem[{\citenamefont{Hult et~al.}(2001)\citenamefont{Hult, Hyldgaard,
  Rossmeisl, and Lundqvist}}]{dft4}
\bibinfo{author}{\bibfnamefont{E.}~\bibnamefont{Hult}},
  \bibinfo{author}{\bibfnamefont{P.}~\bibnamefont{Hyldgaard}},
  \bibinfo{author}{\bibfnamefont{J.}~\bibnamefont{Rossmeisl}},
  \bibnamefont{and} \bibinfo{author}{\bibfnamefont{B.~I.}
  \bibnamefont{Lundqvist}}, \bibinfo{journal}{Phys. Rev. B}
  \textbf{\bibinfo{volume}{64}}, \bibinfo{pages}{195414}
  (\bibinfo{year}{2001}).

\bibitem[{\citenamefont{Bondarev and Lambin}(2004)}]{mbpt1}
\bibinfo{author}{\bibfnamefont{I.~V.} \bibnamefont{Bondarev}} \bibnamefont{and}
  \bibinfo{author}{\bibfnamefont{P.}~\bibnamefont{Lambin}},
  \bibinfo{journal}{Phys. Rev. B} \textbf{\bibinfo{volume}{70}},
  \bibinfo{pages}{035407} (\bibinfo{year}{2004}).

\bibitem[{\citenamefont{Babb et~al.}(2004)\citenamefont{Babb, Klimchitskaya,
  and Mostepanenko}}]{babb}
\bibinfo{author}{\bibfnamefont{J.~F.} \bibnamefont{Babb}},
  \bibinfo{author}{\bibfnamefont{G.~L.} \bibnamefont{Klimchitskaya}},
  \bibnamefont{and} \bibinfo{author}{\bibfnamefont{V.~M.}
  \bibnamefont{Mostepanenko}}, \bibinfo{journal}{Phys. Rev. A}
  \textbf{\bibinfo{volume}{70}}, \bibinfo{pages}{042901}
  (\bibinfo{year}{2004}).

\bibitem[{\citenamefont{Caride et~al.}(2005)\citenamefont{Caride,
  Klimchitskaya, Mostepanenko, and Zanette}}]{caride}
\bibinfo{author}{\bibfnamefont{A.~O.} \bibnamefont{Caride}},
  \bibinfo{author}{\bibfnamefont{G.~L.} \bibnamefont{Klimchitskaya}},
  \bibinfo{author}{\bibfnamefont{V.~M.} \bibnamefont{Mostepanenko}},
  \bibnamefont{and} \bibinfo{author}{\bibfnamefont{S.~I.}
  \bibnamefont{Zanette}}, \bibinfo{journal}{Phys. Rev. A}
  \textbf{\bibinfo{volume}{71}}, \bibinfo{pages}{042901}
  (\bibinfo{year}{2005}).

\bibitem[{\citenamefont{Antezza et~al.}(2004)\citenamefont{Antezza, Pitaevskii,
  and Stringari}}]{lf1}
\bibinfo{author}{\bibfnamefont{M.}~\bibnamefont{Antezza}},
  \bibinfo{author}{\bibfnamefont{L.~P.} \bibnamefont{Pitaevskii}},
  \bibnamefont{and}
  \bibinfo{author}{\bibfnamefont{S.}~\bibnamefont{Stringari}},
  \bibinfo{journal}{Phys. Rev. A} \textbf{\bibinfo{volume}{70}},
  \bibinfo{pages}{053619} (\bibinfo{year}{2004}).

\bibitem[{\citenamefont{Bihmann and Welsch}(2007)}]{lf2}
\bibinfo{author}{\bibfnamefont{S.~Y.} \bibnamefont{Bihmann}} \bibnamefont{and}
  \bibinfo{author}{\bibfnamefont{D.~G.} \bibnamefont{Welsch}},
  \bibinfo{journal}{Prog. Quantum Electron.} \textbf{\bibinfo{volume}{31}},
  \bibinfo{pages}{51} (\bibinfo{year}{2007}).

\bibitem[{\citenamefont{Churkin et~al.}(2011)\citenamefont{Churkin, Fedortsov,
  Klimchitskaya, and Yurova}}]{nano1}
\bibinfo{author}{\bibfnamefont{Y.~V.} \bibnamefont{Churkin}},
  \bibinfo{author}{\bibfnamefont{A.~B.} \bibnamefont{Fedortsov}},
  \bibinfo{author}{\bibfnamefont{G.~L.} \bibnamefont{Klimchitskaya}},
  \bibnamefont{and} \bibinfo{author}{\bibfnamefont{V.~A.}
  \bibnamefont{Yurova}}, \bibinfo{journal}{International Journal of Modern
  Physics: Conference Series} \textbf{\bibinfo{volume}{3}},
  \bibinfo{pages}{555} (\bibinfo{year}{2011}).

\bibitem[{\citenamefont{Chaichian et~al.}(2012)\citenamefont{Chaichian,
  Klimchitskaya, Mostepanenko, and Tureanu}}]{dirac-hydro2}
\bibinfo{author}{\bibfnamefont{M.}~\bibnamefont{Chaichian}},
  \bibinfo{author}{\bibfnamefont{G.~L.} \bibnamefont{Klimchitskaya}},
  \bibinfo{author}{\bibfnamefont{V.~M.} \bibnamefont{Mostepanenko}},
  \bibnamefont{and} \bibinfo{author}{\bibfnamefont{A.}~\bibnamefont{Tureanu}},
  \bibinfo{journal}{Phys. Rev. A} \textbf{\bibinfo{volume}{86}},
  \bibinfo{pages}{012515} (\bibinfo{year}{2012}).

\bibitem[{\citenamefont{Bordag}(2006)}]{bordag}
\bibinfo{author}{\bibfnamefont{M.}~\bibnamefont{Bordag}}, \bibinfo{journal}{J.
  Phys. A} \textbf{\bibinfo{volume}{39}}, \bibinfo{pages}{6173}
  (\bibinfo{year}{2006}).

\bibitem[{\citenamefont{Jiang et~al.}(2009)\citenamefont{Jiang, Du, and
  Dai}}]{metal1}
\bibinfo{author}{\bibfnamefont{D.}~\bibnamefont{Jiang}},
  \bibinfo{author}{\bibfnamefont{M.-H.} \bibnamefont{Du}}, \bibnamefont{and}
  \bibinfo{author}{\bibfnamefont{S.}~\bibnamefont{Dai}}, \bibinfo{journal}{J.
  Chem. Phys.} \textbf{\bibinfo{volume}{130}}, \bibinfo{pages}{074705}
  (\bibinfo{year}{2009}).

\bibitem[{\citenamefont{Wintterlin and Bocquet}(2009)}]{metal2}
\bibinfo{author}{\bibfnamefont{J.}~\bibnamefont{Wintterlin}} \bibnamefont{and}
  \bibinfo{author}{\bibfnamefont{M.~L.} \bibnamefont{Bocquet}},
  \bibinfo{journal}{Surface Science} \textbf{\bibinfo{volume}{603}},
  \bibinfo{pages}{1841} (\bibinfo{year}{2009}).

\bibitem[{\citenamefont{Bezerra et~al.}(2011)\citenamefont{Bezerra, de~Mello,
  Klimchitskaya1, Mostepanenko, and Saharian}}]{cylinder}
\bibinfo{author}{\bibfnamefont{V.~B.} \bibnamefont{Bezerra}},
  \bibinfo{author}{\bibfnamefont{E.~R.~B.} \bibnamefont{de~Mello}},
  \bibinfo{author}{\bibfnamefont{G.~L.} \bibnamefont{Klimchitskaya1}},
  \bibinfo{author}{\bibfnamefont{V.~M.} \bibnamefont{Mostepanenko}},
  \bibnamefont{and} \bibinfo{author}{\bibfnamefont{A.~A.}
  \bibnamefont{Saharian}}, \bibinfo{journal}{Eur. Phys. J. C}
  \textbf{\bibinfo{volume}{71}}, \bibinfo{pages}{1614} (\bibinfo{year}{2011}).

\bibitem[{\citenamefont{Bordag et~al.}(2009{\natexlab{a}})\citenamefont{Bordag,
  Klimchitskaya, Mohideen, and Mostepanenko}}]{harjeet1}
\bibinfo{author}{\bibfnamefont{M.}~\bibnamefont{Bordag}},
  \bibinfo{author}{\bibfnamefont{G.}~\bibnamefont{Klimchitskaya}},
  \bibinfo{author}{\bibfnamefont{U.}~\bibnamefont{Mohideen}}, \bibnamefont{and}
  \bibinfo{author}{\bibfnamefont{V.}~\bibnamefont{Mostepanenko}},
  \emph{\bibinfo{title}{Advances in the Casimir Effect}}
  (\bibinfo{publisher}{Oxford University Press, Oxford},
  \bibinfo{year}{2009}{\natexlab{a}}).

\bibitem[{\citenamefont{Lifshitz and Pitaevskii}(1980)}]{lifshitzbook}
\bibinfo{author}{\bibfnamefont{E.~M.} \bibnamefont{Lifshitz}} \bibnamefont{and}
  \bibinfo{author}{\bibfnamefont{L.~P.} \bibnamefont{Pitaevskii}},
  \emph{\bibinfo{title}{Statistical Physics}} (\bibinfo{publisher}{Pergamon
  Press, Oxford}, \bibinfo{address}{Oxford}, \bibinfo{year}{1980}).

\bibitem[{\citenamefont{Klimchitskaya et~al.}(2009)\citenamefont{Klimchitskaya,
  Mohideen, and Mostepanenko}}]{klim}
\bibinfo{author}{\bibfnamefont{G.~L.} \bibnamefont{Klimchitskaya}},
  \bibinfo{author}{\bibfnamefont{U.}~\bibnamefont{Mohideen}}, \bibnamefont{and}
  \bibinfo{author}{\bibfnamefont{V.~M.} \bibnamefont{Mostepanenko}},
  \bibinfo{journal}{Rev. Mod. Phys.} \textbf{\bibinfo{volume}{81}},
  \bibinfo{pages}{1827} (\bibinfo{year}{2009}).

\bibitem[{\citenamefont{Mahanty and Ninham}(1976)}]{mahanty}
\bibinfo{author}{\bibfnamefont{J.}~\bibnamefont{Mahanty}} \bibnamefont{and}
  \bibinfo{author}{\bibfnamefont{B.~W.} \bibnamefont{Ninham}},
  \emph{\bibinfo{title}{Dispersion Forces}} (\bibinfo{publisher}{Academic
  Press}, \bibinfo{address}{New York}, \bibinfo{year}{1976}).

\bibitem[{\citenamefont{Parsegian}(2005)}]{pars}
\bibinfo{author}{\bibfnamefont{V.~A.} \bibnamefont{Parsegian}},
  \emph{\bibinfo{title}{Van der Waals Forces: A Handbook for Biologists,
  Chemists, Engineers, and Physicists}} (\bibinfo{publisher}{Cambridge
  University Press, Cambridge}, \bibinfo{year}{2005}).

\bibitem[{\citenamefont{Arora and Sahoo}(2012)}]{arora-sahoo1}
\bibinfo{author}{\bibfnamefont{B.}~\bibnamefont{Arora}} \bibnamefont{and}
  \bibinfo{author}{\bibfnamefont{B.~K.} \bibnamefont{Sahoo}},
  \bibinfo{journal}{Phys. Rev. A} \textbf{\bibinfo{volume}{86}},
  \bibinfo{pages}{033416} (\bibinfo{year}{2012}).

\bibitem[{\citenamefont{Sahoo and Arora}(2013)}]{arora-sahoo2}
\bibinfo{author}{\bibfnamefont{B.~K.} \bibnamefont{Sahoo}} \bibnamefont{and}
  \bibinfo{author}{\bibfnamefont{B.}~\bibnamefont{Arora}},
  \bibinfo{journal}{Phys. Rev. A} \textbf{\bibinfo{volume}{87}},
  \bibinfo{pages}{023402} (\bibinfo{year}{2013}).

\bibitem[{\citenamefont{Barton}(2004)}]{barton}
\bibinfo{author}{\bibfnamefont{G.}~\bibnamefont{Barton}}, \bibinfo{journal}{J.
  Phys. A} \textbf{\bibinfo{volume}{37}}, \bibinfo{pages}{1011}
  (\bibinfo{year}{2004}).

\bibitem[{\citenamefont{Barton}(2005)}]{barton1}
\bibinfo{author}{\bibfnamefont{G.}~\bibnamefont{Barton}}, \bibinfo{journal}{J.
  Phys. A} \textbf{\bibinfo{volume}{38}}, \bibinfo{pages}{2997}
  (\bibinfo{year}{2005}).

\bibitem[{\citenamefont{Bordag et~al.}(2001)\citenamefont{Bordag, Mohideen, and
  Mostepanenko}}]{harjeet2}
\bibinfo{author}{\bibfnamefont{M.}~\bibnamefont{Bordag}},
  \bibinfo{author}{\bibfnamefont{U.}~\bibnamefont{Mohideen}}, \bibnamefont{and}
  \bibinfo{author}{\bibfnamefont{V.~M.} \bibnamefont{Mostepanenko}},
  \bibinfo{journal}{Phys. Rep.} \textbf{\bibinfo{volume}{353}},
  \bibinfo{pages}{1} (\bibinfo{year}{2001}).

\bibitem[{\citenamefont{Bordag et~al.}(2006)\citenamefont{Bordag, Geyer,
  Klimchitskaya, and Mostepanenko}}]{bordag1}
\bibinfo{author}{\bibfnamefont{M.}~\bibnamefont{Bordag}},
  \bibinfo{author}{\bibfnamefont{B.}~\bibnamefont{Geyer}},
  \bibinfo{author}{\bibfnamefont{G.~L.} \bibnamefont{Klimchitskaya}},
  \bibnamefont{and} \bibinfo{author}{\bibfnamefont{V.~M.}
  \bibnamefont{Mostepanenko}}, \bibinfo{journal}{Phys. Rev. B}
  \textbf{\bibinfo{volume}{74}}, \bibinfo{pages}{205431}
  (\bibinfo{year}{2006}).

\bibitem[{\citenamefont{Blagov et~al.}(2007)\citenamefont{Blagov,
  Klimchitskaya, and Mostepanenko}}]{blagov}
\bibinfo{author}{\bibfnamefont{E.~V.} \bibnamefont{Blagov}},
  \bibinfo{author}{\bibfnamefont{G.~L.} \bibnamefont{Klimchitskaya}},
  \bibnamefont{and} \bibinfo{author}{\bibfnamefont{V.~M.}
  \bibnamefont{Mostepanenko}}, \bibinfo{journal}{Phys. Rev. B}
  \textbf{\bibinfo{volume}{75}}, \bibinfo{pages}{235413}
  (\bibinfo{year}{2007}).

\bibitem[{\citenamefont{Bordag et~al.}(2009{\natexlab{b}})\citenamefont{Bordag,
  Fialkovsky, Gitman, and Vassilevich}}]{bordag2}
\bibinfo{author}{\bibfnamefont{M.}~\bibnamefont{Bordag}},
  \bibinfo{author}{\bibfnamefont{I.~V.} \bibnamefont{Fialkovsky}},
  \bibinfo{author}{\bibfnamefont{D.~M.} \bibnamefont{Gitman}},
  \bibnamefont{and} \bibinfo{author}{\bibfnamefont{D.~V.}
  \bibnamefont{Vassilevich}}, \bibinfo{journal}{Phys. Rev. B}
  \textbf{\bibinfo{volume}{80}}, \bibinfo{pages}{245406}
  (\bibinfo{year}{2009}{\natexlab{b}}).

\bibitem[{\citenamefont{Lifshitz}(1955)}]{lifshitz1}
\bibinfo{author}{\bibfnamefont{E.~M.} \bibnamefont{Lifshitz}},
  \bibinfo{journal}{Zh. Exsp. Toer. Fiz.} \textbf{\bibinfo{volume}{29}},
  \bibinfo{pages}{94} (\bibinfo{year}{1955}).

\bibitem[{\citenamefont{Blagov et~al.}(2005)\citenamefont{Blagov,
  Klimchitskaya, and Mostepanenko}}]{som1}
\bibinfo{author}{\bibfnamefont{E.~V.} \bibnamefont{Blagov}},
  \bibinfo{author}{\bibfnamefont{G.~L.} \bibnamefont{Klimchitskaya}},
  \bibnamefont{and} \bibinfo{author}{\bibfnamefont{V.~M.}
  \bibnamefont{Mostepanenko}}, \bibinfo{journal}{Phys. rev. B}
  \textbf{\bibinfo{volume}{71}}, \bibinfo{pages}{235401}
  (\bibinfo{year}{2005}).

\bibitem[{\citenamefont{Casimir and Polder}(1948)}]{casimir}
\bibinfo{author}{\bibfnamefont{H.~B.~G.} \bibnamefont{Casimir}}
  \bibnamefont{and} \bibinfo{author}{\bibfnamefont{D.}~\bibnamefont{Polder}},
  \bibinfo{journal}{Phys. Rev.} \textbf{\bibinfo{volume}{73}},
  \bibinfo{pages}{360} (\bibinfo{year}{1948}).

\bibitem[{\citenamefont{Safronova et~al.}(2012)\citenamefont{Safronova,
  Safronova, and Clark}}]{safronova-li}
\bibinfo{author}{\bibfnamefont{M.~S.} \bibnamefont{Safronova}},
  \bibinfo{author}{\bibfnamefont{U.~I.} \bibnamefont{Safronova}},
  \bibnamefont{and} \bibinfo{author}{\bibfnamefont{C.~W.} \bibnamefont{Clark}},
  \bibinfo{journal}{Phys. Rev. A} \textbf{\bibinfo{volume}{86}},
  \bibinfo{pages}{042505} (\bibinfo{year}{2012}).

\bibitem[{\citenamefont{Arora et~al.}(2007)\citenamefont{Arora, Safronova, and
  Clark}}]{arora1}
\bibinfo{author}{\bibfnamefont{B.}~\bibnamefont{Arora}},
  \bibinfo{author}{\bibfnamefont{M.~S.} \bibnamefont{Safronova}},
  \bibnamefont{and} \bibinfo{author}{\bibfnamefont{C.~W.} \bibnamefont{Clark}},
  \bibinfo{journal}{Phys. Rev. A} \textbf{\bibinfo{volume}{76}},
  \bibinfo{pages}{052509} (\bibinfo{year}{2007}).

\bibitem[{\citenamefont{Derevianko et~al.}(1999)\citenamefont{Derevianko,
  Johnson, Safronova, and Babb}}]{pol-andrei}
\bibinfo{author}{\bibfnamefont{A.}~\bibnamefont{Derevianko}},
  \bibinfo{author}{\bibfnamefont{W.~R.} \bibnamefont{Johnson}},
  \bibinfo{author}{\bibfnamefont{M.~S.} \bibnamefont{Safronova}},
  \bibnamefont{and} \bibinfo{author}{\bibfnamefont{J.~F.} \bibnamefont{Babb}},
  \bibinfo{journal}{Phys.\ Rev.\ Lett.} \textbf{\bibinfo{volume}{82}},
  \bibinfo{pages}{3589} (\bibinfo{year}{1999}).

\bibitem[{\citenamefont{Arora et~al.}(2012)\citenamefont{Arora, Nandy, and
  Sahoo}}]{arora-sahoo3}
\bibinfo{author}{\bibfnamefont{B.}~\bibnamefont{Arora}},
  \bibinfo{author}{\bibfnamefont{D.~K.} \bibnamefont{Nandy}}, \bibnamefont{and}
  \bibinfo{author}{\bibfnamefont{B.~K.} \bibnamefont{Sahoo}},
  \bibinfo{journal}{Phys. Rev. A} \textbf{\bibinfo{volume}{85}},
  \bibinfo{pages}{012506} (\bibinfo{year}{2012}).

\bibitem[{\citenamefont{Volz and Schmoranzer}(1996)}]{volz}
\bibinfo{author}{\bibfnamefont{U.}~\bibnamefont{Volz}} \bibnamefont{and}
  \bibinfo{author}{\bibfnamefont{H.}~\bibnamefont{Schmoranzer}},
  \bibinfo{journal}{Phys. Scr. T} \textbf{\bibinfo{volume}{65}},
  \bibinfo{pages}{48} (\bibinfo{year}{1996}).

\bibitem[{\citenamefont{Miffre et~al.}(2006)\citenamefont{Miffre, Jacquest,
  Buchner, Trenec, and Vigue}}]{li-exp}
\bibinfo{author}{\bibfnamefont{A.}~\bibnamefont{Miffre}},
  \bibinfo{author}{\bibfnamefont{M.}~\bibnamefont{Jacquest}},
  \bibinfo{author}{\bibfnamefont{M.}~\bibnamefont{Buchner}},
  \bibinfo{author}{\bibfnamefont{G.}~\bibnamefont{Trenec}}, \bibnamefont{and}
  \bibinfo{author}{\bibfnamefont{J.}~\bibnamefont{Vigue}},
  \bibinfo{journal}{Eur. Phys. J. D} \textbf{\bibinfo{volume}{38}},
  \bibinfo{pages}{353} (\bibinfo{year}{2006}).

\bibitem[{\citenamefont{Ekstrom
  et~al.}(1995{\natexlab{a}})\citenamefont{Ekstrom, Schmiedmayer, Chapman,
  Hammond, and Pritchard}}]{na-exp}
\bibinfo{author}{\bibfnamefont{C.~R.} \bibnamefont{Ekstrom}},
  \bibinfo{author}{\bibfnamefont{J.}~\bibnamefont{Schmiedmayer}},
  \bibinfo{author}{\bibfnamefont{M.~S.} \bibnamefont{Chapman}},
  \bibinfo{author}{\bibfnamefont{T.~D.} \bibnamefont{Hammond}},
  \bibnamefont{and} \bibinfo{author}{\bibfnamefont{D.~E.}
  \bibnamefont{Pritchard}}, \bibinfo{journal}{Phys. Rev. A}
  \textbf{\bibinfo{volume}{51}}, \bibinfo{pages}{3883}
  (\bibinfo{year}{1995}{\natexlab{a}}).

\bibitem[{\citenamefont{Holmgren et~al.}(2010)\citenamefont{Holmgren, Revelle,
  Lonij, and Cronin}}]{k-exp}
\bibinfo{author}{\bibfnamefont{W.~F.} \bibnamefont{Holmgren}},
  \bibinfo{author}{\bibfnamefont{M.~C.} \bibnamefont{Revelle}},
  \bibinfo{author}{\bibfnamefont{V.~P.~A.} \bibnamefont{Lonij}},
  \bibnamefont{and} \bibinfo{author}{\bibfnamefont{A.~D.}
  \bibnamefont{Cronin}}, \bibinfo{journal}{Phys. Rev. A}
  \textbf{\bibinfo{volume}{81}}, \bibinfo{pages}{053607}
  (\bibinfo{year}{2010}).

\bibitem[{\citenamefont{W.~F.~Holmgren and Cronin}(2010)}]{rb-exp}
\bibinfo{author}{\bibfnamefont{V.~P. A.~L.} \bibnamefont{W.~F.~Holmgren},
  \bibfnamefont{M.~C.~Revelle}} \bibnamefont{and}
  \bibinfo{author}{\bibfnamefont{A.~D.} \bibnamefont{Cronin}},
  \bibinfo{journal}{Phys. Rev. A} \textbf{\bibinfo{volume}{81}},
  \bibinfo{pages}{053607} (\bibinfo{year}{2010}).

\bibitem[{\citenamefont{Ekstrom
  et~al.}(1995{\natexlab{b}})\citenamefont{Ekstrom, Schmiedmayer, Chapman,
  Hammond, and Pritchard}}]{13}
\bibinfo{author}{\bibfnamefont{C.}~\bibnamefont{Ekstrom}},
  \bibinfo{author}{\bibfnamefont{J.}~\bibnamefont{Schmiedmayer}},
  \bibinfo{author}{\bibfnamefont{M.}~\bibnamefont{Chapman}},
  \bibinfo{author}{\bibfnamefont{T.}~\bibnamefont{Hammond}}, \bibnamefont{and}
  \bibinfo{author}{\bibfnamefont{D.}~\bibnamefont{Pritchard}},
  \bibinfo{journal}{Phys. Rev. A} \textbf{\bibinfo{volume}{51}},
  \bibinfo{pages}{3883} (\bibinfo{year}{1995}{\natexlab{b}}).

\bibitem[{\citenamefont{Molof et~al.}(1974)\citenamefont{Molof, Schwartz,
  Miller, and Bederson}}]{14}
\bibinfo{author}{\bibfnamefont{R.}~\bibnamefont{Molof}},
  \bibinfo{author}{\bibfnamefont{H.}~\bibnamefont{Schwartz}},
  \bibinfo{author}{\bibfnamefont{T.}~\bibnamefont{Miller}}, \bibnamefont{and}
  \bibinfo{author}{\bibfnamefont{B.}~\bibnamefont{Bederson}},
  \bibinfo{journal}{Phys. Rev. A} \textbf{\bibinfo{volume}{10}},
  \bibinfo{pages}{1131} (\bibinfo{year}{1974}).

\bibitem[{\citenamefont{Hall and Zorn}(1971)}]{15}
\bibinfo{author}{\bibfnamefont{W.~D.} \bibnamefont{Hall}} \bibnamefont{and}
  \bibinfo{author}{\bibfnamefont{J.~C.} \bibnamefont{Zorn}},
  \bibinfo{journal}{Rhys. Rev. A} \textbf{\bibinfo{volume}{10}},
  \bibinfo{pages}{1141} (\bibinfo{year}{1971}).

\bibitem[{\citenamefont{Mukherjee et~al.}(2009)\citenamefont{Mukherjee, Sahoo,
  Nataraj, and Das}}]{mukherjee}
\bibinfo{author}{\bibfnamefont{D.}~\bibnamefont{Mukherjee}},
  \bibinfo{author}{\bibfnamefont{B.~K.} \bibnamefont{Sahoo}},
  \bibinfo{author}{\bibfnamefont{H.~S.} \bibnamefont{Nataraj}},
  \bibnamefont{and} \bibinfo{author}{\bibfnamefont{B.~P.} \bibnamefont{Das}},
  \bibinfo{journal}{J. Phys. Chem. A} \textbf{\bibinfo{volume}{113}},
  \bibinfo{pages}{12549} (\bibinfo{year}{2009}).

\bibitem[{\citenamefont{Sahoo et~al.}(2004)\citenamefont{Sahoo, Majumder,
  Chaudhuri, Das, and Mukhrejee}}]{sahoo2}
\bibinfo{author}{\bibfnamefont{B.~K.} \bibnamefont{Sahoo}},
  \bibinfo{author}{\bibfnamefont{S.}~\bibnamefont{Majumder}},
  \bibinfo{author}{\bibfnamefont{R.~K.} \bibnamefont{Chaudhuri}},
  \bibinfo{author}{\bibfnamefont{B.~P.} \bibnamefont{Das}}, \bibnamefont{and}
  \bibinfo{author}{\bibfnamefont{D.}~\bibnamefont{Mukhrejee}},
  \bibinfo{journal}{J. Phys. B} \textbf{\bibinfo{volume}{37}},
  \bibinfo{pages}{3409} (\bibinfo{year}{2004}).

\bibitem[{\citenamefont{Arora and Sahoo}(2013)}]{bindiya3}
\bibinfo{author}{\bibfnamefont{B.}~\bibnamefont{Arora}} \bibnamefont{and}
  \bibinfo{author}{\bibfnamefont{B.~K.} \bibnamefont{Sahoo}},
  \emph{\bibinfo{title}{Van der waals coefficients for the alkali-metal atoms
  in the material mediums}} (\bibinfo{year}{2013}),
  \bibinfo{note}{arXiv:1309.4897}.

\bibitem[{\citenamefont{Rodríguez-Manzo
  et~al.}(2010)\citenamefont{Rodríguez-Manzo, Cretu, and Banhart}}]{trapping}
\bibinfo{author}{\bibfnamefont{J.~A.} \bibnamefont{Rodríguez-Manzo}},
  \bibinfo{author}{\bibfnamefont{O.}~\bibnamefont{Cretu}}, \bibnamefont{and}
  \bibinfo{author}{\bibfnamefont{F.}~\bibnamefont{Banhart}},
  \bibinfo{journal}{ACS nano} \textbf{\bibinfo{volume}{4}},
  \bibinfo{pages}{3422} (\bibinfo{year}{2010}).

\bibitem[{\citenamefont{Petrov et~al.}(2009)\citenamefont{Petrov, Machluf,
  Younis, Macaluso, David, Hadad, Japha, Keil, Joselevich, and
  Folman}}]{trapping1}
\bibinfo{author}{\bibfnamefont{P.~G.} \bibnamefont{Petrov}},
  \bibinfo{author}{\bibfnamefont{S.}~\bibnamefont{Machluf}},
  \bibinfo{author}{\bibfnamefont{S.}~\bibnamefont{Younis}},
  \bibinfo{author}{\bibfnamefont{R.}~\bibnamefont{Macaluso}},
  \bibinfo{author}{\bibfnamefont{T.}~\bibnamefont{David}},
  \bibinfo{author}{\bibfnamefont{B.}~\bibnamefont{Hadad}},
  \bibinfo{author}{\bibfnamefont{Y.}~\bibnamefont{Japha}},
  \bibinfo{author}{\bibfnamefont{M.}~\bibnamefont{Keil}},
  \bibinfo{author}{\bibfnamefont{E.}~\bibnamefont{Joselevich}},
  \bibnamefont{and} \bibinfo{author}{\bibfnamefont{R.}~\bibnamefont{Folman}},
  \bibinfo{journal}{Phys. Rev. A} \textbf{\bibinfo{volume}{79}},
  \bibinfo{pages}{043403} (\bibinfo{year}{2009}).

\bibitem[{\citenamefont{Rodríguez-Manzo and Banhart}(2009)}]{trapping2}
\bibinfo{author}{\bibfnamefont{J.~A.} \bibnamefont{Rodríguez-Manzo}}
  \bibnamefont{and} \bibinfo{author}{\bibfnamefont{F.}~\bibnamefont{Banhart}},
  \bibinfo{journal}{Nano Lett.} \textbf{\bibinfo{volume}{9}},
  \bibinfo{pages}{2285} (\bibinfo{year}{2009}).

\bibitem[{\citenamefont{Cretu et~al.}(2010)\citenamefont{Cretu, Krasheninnikov,
  Rodriguez-Manzo, Sun, Nieminen, , and Banhart}}]{trapping3}
\bibinfo{author}{\bibfnamefont{O.}~\bibnamefont{Cretu}},
  \bibinfo{author}{\bibfnamefont{A.~V.} \bibnamefont{Krasheninnikov}},
  \bibinfo{author}{\bibfnamefont{J.~A.} \bibnamefont{Rodriguez-Manzo}},
  \bibinfo{author}{\bibfnamefont{L.}~\bibnamefont{Sun}},
  \bibinfo{author}{\bibfnamefont{R.~M.} \bibnamefont{Nieminen}}, ,
  \bibnamefont{and} \bibinfo{author}{\bibfnamefont{F.}~\bibnamefont{Banhart}},
  \bibinfo{journal}{Phys. Rev. Lett.} \textbf{\bibinfo{volume}{105}},
  \bibinfo{pages}{196102} (\bibinfo{year}{2010}).

\bibitem[{\citenamefont{Krasheninnikov and Nieminen}(2011)}]{trapping4}
\bibinfo{author}{\bibfnamefont{A.~V.} \bibnamefont{Krasheninnikov}}
  \bibnamefont{and} \bibinfo{author}{\bibfnamefont{R.~M.}
  \bibnamefont{Nieminen}}, \bibinfo{journal}{Theor. Chem. Acc.}
  \textbf{\bibinfo{volume}{129}}, \bibinfo{pages}{625} (\bibinfo{year}{2011}).

\bibitem[{\citenamefont{Tang et~al.}(2011)\citenamefont{Tang, Yang, and
  Dai}}]{trapping5}
\bibinfo{author}{\bibfnamefont{Y.}~\bibnamefont{Tang}},
  \bibinfo{author}{\bibfnamefont{Z.}~\bibnamefont{Yang}}, \bibnamefont{and}
  \bibinfo{author}{\bibfnamefont{X.}~\bibnamefont{Dai}}, \bibinfo{journal}{J.
  Chem. Phys.} \textbf{\bibinfo{volume}{135}}, \bibinfo{pages}{224704}
  (\bibinfo{year}{2011}).

\end{thebibliography}

\end{document}